\documentclass[letterpaper,twocolumn,10pt]{article}
\usepackage{usenix-2020-09}

\usepackage{tikz}

\usepackage{pifont}
\usepackage{amsmath}
\usepackage{totpages}
\usepackage{comment}
\usepackage{balance}

\usepackage{main_macros}

\begin{document}

\date{}


\title{Isolation Mechanisms for High-Speed Packet-Processing Pipelines}





\ifarxiv
\author{
{\rm Tao Wang$^\dagger$}
\and
{\rm Xiangrui Yang$^\ddagger$\thanks{Work done at Queen Mary University of London}}
\and 
{\rm Gianni Antichi$^{\star\star}$}
\and
{\rm Anirudh Sivaraman$^\dagger$}
\and
{\rm Aurojit Panda$^\dagger$}
\and 
{$^\dagger$New York University~~~$^\ddagger$National University of Defense Technology}
\and
{$^{\star\star}$Queen Mary University of London}
}
\else 
\fi

\maketitle

\begin{abstract}
Data-plane programmability is now mainstream. As we find more use cases, deployments need to be able to run multiple packet-processing modules in a single device. These are likely to be developed by independent teams, either within the same organization or from multiple organizations. Therefore, we need isolation mechanisms to ensure that modules on the same device do not interfere with each other.

This paper presents \sysname, an extension of the Reconfigurable Match Tables (RMT) pipeline that enforces isolation between different packet-processing modules. \sysname is comprised of a set of lightweight hardware primitives and an extension to the open source P4-16 reference compiler that act in conjunction to meet this goal. We have prototyped \sysname on two FPGA platforms (NetFPGA and Corundum). We show that our design provides isolation, and allows new modules to be loaded without impacting the ones already running. Finally, we demonstrate the feasibility of implementing \sysname on ASICs by using the FreePDK45nm technology library and the Synopsys DC synthesis software, showing that our design meets timing at a 1 GHz clock frequency and needs approximately 6\% additional chip area. We have open sourced the code for \sysname's hardware and software at \gitrepo.

\end{abstract}


\section{Introduction}
\label{sec:intro}
Programmable network devices in the form of programmable switches~\cite{tofino, mellanox_spectrum, trident} and smart network interface cards (SmartNICs)~\cite{mellanox_bluefield, azure_smartnic, liquidio} are becoming commodity. Such devices allow the network infrastructure to provide its users additional services beyond packet forwarding, e.g., congestion control~\cite{rcp, hpcc}, measurement~\cite{sonata}, load balancing~\cite{hula}, in-network caches~\cite{netcache}, and machine learning~\cite{switchml}.

As network programmability matures, a single device will have to concurrently support {\em multiple} independently developed modules. This is the case for {\em networks in the public cloud} where tenants can provide packet-processing modules that are installed and run on the cloud provider's devices. Another example is when different teams in an organization write different modules, e.g., an in-networking caching module and a telemetry module.

{\em Isolation} is required to safely run multiple modules on a single device. Several prior projects have observed this need and proposed solutions targeting multicore network processors~\cite{fairnic, ipipe}, FPGA-based packet processors~\cite{mtpsa, virtp4, online_reconf_switch, p4visor}, and software switches~\cite{hyper4, hyperv}. However, thus far, high-speed pipelines such as RMT that are used in switch and NIC ASICs provide only limited support for isolation. For instance, the Tofino programmable switch ASIC~\cite{tofino} provides mechanisms to share stateful memory across modules but cannot share other resources, e.g., match-action tables~\cite{multitenant_hotcloud}.

\begin{figure}[!t]
    \centering
    \includegraphics[width=0.9\linewidth]{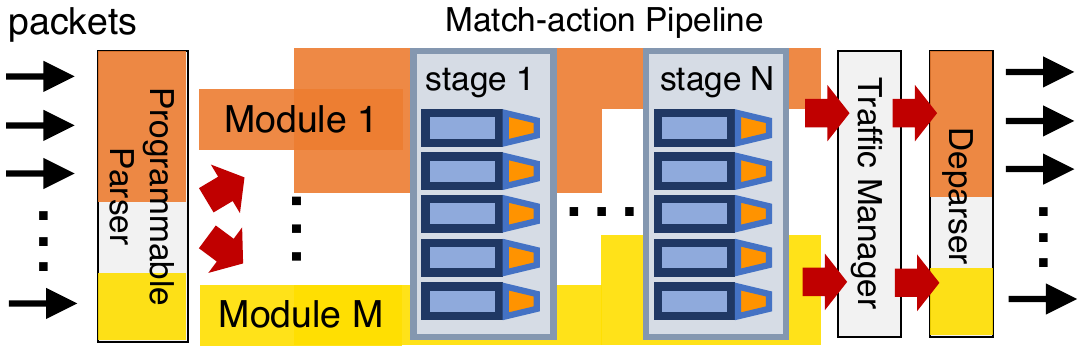}
    \caption{The RMT architecture~\cite{rmt} typically consists of a programmable parser/deparser, match-action pipeline and traffic manager. \sysname provides isolation between RMT modules.
    In the figure we show resources allocated to \textcolor{anotherorange}{module 1} and \textcolor{goldenyellow}{module m} by shading them in the appropriate color.\label{fig:rmt}}
\end{figure}

Our goal with this paper is to lay out the requirements for isolation mechanisms on the RMT architecture that are applicable to all resources and then to design lightweight mechanisms that meet these requirements. As presented in Figure~\ref{fig:rmt}, the desired isolation mechanisms should guarantee that multiple modules can be allocated to different resources, and process packets in parallel without impacting each other.
In brief (\S\ref{ssec:requirements} elaborates), we seek isolation mechanisms that ensure that (1) one module's behavior (input, output, and internal state) is unaffected by another module; (2) one module can not affect another's throughput and/or latency; and (3) one module can not access RMT pipeline resources belonging to another. Given the high performance requirements of RMT, we also seek mechanisms that are lightweight. Finally, the isolation mechanism should ensure that one module can be updated without disturbing any other modules and that the update process itself is quick.

The RMT architecture poses unique challenges for isolation because its pipeline design means that neither an OS nor a hypervisor can be used to enforce isolation.\footnote{An OS does run on the network device's control CPU, allowing isolation in the control plane. Our focus, instead, is on isolation in the data plane.} This is because RMT is a {\em dataflow} or spatial hardware architecture~\cite{arvind_dataflow, dennis_dataflow} with a set of instructions units continuously processing data (packets). This is in contrast to the Von Neumann architecture found on processors~\cite{von_neumann}, where a program counter decides what instruction to execute next. As such, an RMT pipeline is closer in its hardware architecture to an FPGA or a CGRA~\cite{cgra} than a processor.
This difference in architecture has important implications for isolation. The Von Neumann architecture supports a {\em time-sharing} approach to isolation (in the form of an OS/hypervisor) that runs different modules on the CPU successively by changing the program counter to point to the next instruction of the next module. We instead use {\em space-partitioning} to divide up the RMT pipeline's resources (e.g., match-action tables) across different modules.


Unfortunately, space partitioning is not a viable option for certain RMT resources because there are very few of them to be effectively partitioned across modules (e.g., match key extraction units (\S\ref{subsec:hardware})). For such resources, we add additional hardware primitives in the form of small tables that store module-specific configurations for these resources. As a packet progresses through the pipeline, the packet's module identifier is used as an index into these tables to extract module-specific configurations before processing the packet according to the just extracted configuration. These primitives are similar to the use of {\em overlays}~\cite{ostep, denning_overlay} in embedded systems~\cite{nap_avionics_report, arm_developer} and earlier PCs~\cite{commodore}. They effectively allow us to bring in different configurations for the same RMT resource, in response to different packets from different modules.

Based on the ideas of space partitioning and overlays, we build a system, \sysname, for isolation on RMT pipelines. Specifically, \sysname makes the following contributions:
\begin{CompactEnumerate}
\item The use of space partitioning and overlays as techniques to achieve isolation when sharing an RMT pipeline across multiple modules.
\item A hardware design for an RMT pipeline that employs these techniques.
\item An implementation on 2 open-source FPGA platforms: the NetFPGA switch~\cite{netfpga} and Corundum NIC~\cite{corundum}.
\item A compiler based on the open-source P4-16 compiler~\cite{p4c} that supports multiple modules running on RMT, along with a system-level module to provide basic services (e.g., routing, multicast) to other modules.
\item An evaluation of \sysname using 8 modules---based on tutorial P4 programs, and the NetCache~\cite{netcache} and NetChain~\cite{netchain} research projects---showing that \sysname meets our isolation requirements.
\item An ASIC analysis of the \sysname, which shows that our design can meet timing at 1 GHz (comparable to current programmable ASICs) with modest additional area relative to a baseline RMT design.
\end{CompactEnumerate}

Overall, we find that \sysname adds modest overhead to an existing RMT pipeline in both FPGA and ASIC implementations (\S\ref{sec:evaluation}). Our main takeaway is that a small number of simple additions to RMT along with changes to the RMT compiler can provide inter-module isolation for a high-speed packet-processing pipeline. We have made \sysname's hardware design and software available under an open-source license at \gitrepo to enable further research into isolation mechanisms for high-speed pipelines. 
\section{The case for isolation}
\label{sec:motivation}
\label{ssec:usecases}
A single network device might host a measurement module~\cite{sonata}, a forwarding module~\cite{dc_p4}, an in-network caching~\cite{netcache} module, and an in-network machine-learning module~\cite{switchml}---each written by a different team in the same organization. It is important to isolate these modules from each other. This would prevent bugs in measurement, in-network caching, and in-network ML from causing network downtime. It would also ensure that memory for measuring per-flow stats~\cite{li16-flowradar} is separated from memory for routing tables, e.g., a sudden arrival of many new flows does not cause cached routes to be evicted from the data plane.

The packet-processing modules in question do not even have to be developed by teams in the same organization~\cite{multitenant_hotcloud}. They could belong to different tenants sharing the same public cloud network. This would allow cloud providers to offer network data-plane programmability as a service to their tenants, similar to cloud CPU, GPU, and storage offerings today. Such a capability would allow tenants to customize network devices in the cloud to suit their needs.

\subsection{Requirements for isolation mechanisms}
\label{ssec:requirements}
For the rest of this paper, we will use the term module to refer to a packet-processing program that must be isolated from other such programs, regardless of whether the modules belong to different mutually distrustful tenants or to a single network operator. Importantly, modules can not call each other like functions, but are intended to isolate different pieces of functionality from each other---similar to processes. Based on our use cases above (\S\ref{ssec:usecases}), we want an inter-module isolation mechanisms that meet the requirements below:

\begin{CompactEnumerate}
    \item \textbf{Behavior isolation.} The behavior of one module must not affect the behavior (i.e., input, output, computation and internal state) of another. This would prevent a faulty or malicious module from adversely affecting other modules. Further, one module should not be able to inspect the behavior of another module.
    \item \textbf{Resource isolation.} A switch/NIC pipeline has multiple resources, e.g., static random-access memory (SRAM) for exact matching and ternary content-addressable memory (TCAM) for ternary matching. Each module should be able to access only its assigned subset of the pipeline's resources and no more. It should also be possible to allocate each resource independent of other resources. For example, an in-network caching module may need large amounts of stateful memory~\cite{netcache} for its caches, but a routing module may need significant TCAM for routing tables.
    \item \textbf{Performance isolation.} Each module should stay within its allotted ingress packets per second and bits per second rates. One module's behavior should not affect the throughput and latency of another module.
    \item \textbf{Lightweight.} The isolation mechanisms themselves must have low overhead so that their presence does not significantly degrade the high performance of the underlying network device. In addition, the extra hardware consumed by these mechanisms must be small.
    \item \textbf{Rapid reconfiguration.} If a module is reconfigured with new packet-processing logic, the reconfiguration process should be quick.
    \item \textbf{No disruption.} If a module is reconfigured, it must not disrupt the behavior of other unchanged modules---especially important in a multi-tenant environment~\cite{prism}.
\end{CompactEnumerate}


\subsection{Target setting for \sysname}
\label{ssec:target}
We target both programmable switches and NICs with a programmable packet-processing pipeline based on the RMT pipeline~\cite{rmt}, a common architecture for packet processing for the highest end devices. Other projects have looked at isolation for software switches, multicore network processors, FPGA-based devices, and the Barefoot Tofino switch (without hardware changes). \S\ref{sec:related} compares against them.

An RMT pipeline can be implemented either on an FPGA (e.g., FlowBlaze~\cite{flowblaze}, Lightning NIC~\cite{lnic}, nanoPU~\cite{nanopu}) or an ASIC (e.g., the Tofino~\cite{tofino}, Spectrum~\cite{mellanox_spectrum}, and Trident~\cite{trident} switches; and the Pensando NIC~\cite{pensando_dsc_100}). This pipeline has also been embedded within larger hardware designs (e.g., PANIC~\cite{panic}). \sysname builds on a baseline RMT pipeline to provide isolation between different modules/tenants. A high-speed implementation of \sysname would likely be based on an RMT ASIC. For this paper, we prototype RMT on 2 FPGA-based platforms: the NetFPGA switch~\cite{netfpga} and the Corundum NIC~\cite{corundum}. Our ASIC synthesis results suggest that our lessons generalize to ASICs as well (\S\ref{ssec:perf}). 


\section{Design}
\label{sec:design}


\begin{table}[!t]
\begin{center}
\scriptsize 
    \begin{tabular}{ll}
        \toprule
        \textbf{Applied Mechanism} & \textbf{Targeted Resources} \\
        \midrule
        Space partitioning & Match action table entries, stateful memories \\ 
        \midrule
        Overlays & Parsing actions, key extractors,\\
         & packet header vector (PHV) containers, \\
         & arithmetic logic units (ALUs)  \\
        \bottomrule
    \end{tabular}
\caption{Summary of \sysname's mechanisms.}
\label{tab:machanisms}
\end{center}
\end{table}

In order to meet its performance goals, RMT's pipelined architecture ensures that processing stages never stall, i.e., they can process a packet every clock cycle. The \sysname design aims to preserve this invariant so that isolation does not come at the cost of performance. To maintain this invariant, \sysname's isolation mechanisms cannot reconfigure stages or change table contents between packets. As a result, \sysname provides isolation by spatially partitioning switch resources between packet processing modules.

While spatial partitioning is easy for resources, e.g., match-action tables and stateful memory, that are provisioned so they can be allocated at flow granularity, it is much more challenging for resources such as key extractors (\S\ref{subsec:hardware}) which are generally shared across flows. This is because naive approaches to spatially partitioning such shared resources across packet-processing modules would severely reduce the number of resources available to each packet processing module---and hence the richness of that module.

To see why, consider a case where a key extractor is split between two packet processing modules: in this setting each packet processing module can only use half the key extractor, limiting its key length to half of what it would be able to use were it running on the entire pipeline. This problem is of course further exacerbated as we increase the number of packet processing modules sharing the pipeline.


\sysname addresses this problem using \emph{overlays}: we associate a configuration lookup table with each shared resource in the switch. This lookup table is keyed by the packet processing module's ID and contains the configuration that should be used when processing packets for this module. For example, in the case of the key extractor, the configuration table contains the instructions that the module uses to construct key (\S\ref{subsec:hardware}). Our use of overlays means that we do not need to partition resources including ALUs or PHVs between modules. Instead, the module has exclusive access to all PHVs/ALUs in a stage when processing a packet. Table~\ref{tab:machanisms} summarizes our mechanisms.

To realize \sysname{}, on the software side, we modify an RMT compiler to target a block of resources rather than the entire pipeline. Overlays require new hardware primitives to be added to the RMT pipeline. These hardware primitives are small tables that contain per-module configurations of shared resources. On every packet, these tables are indexed using the packet's module ID to determine the configuration to use for that packet at that resource. An incremental deployment pathway for \sysname would be to only modify an RMT compiler (e.g., Tofino's compiler) to implement space partitioning without investing in new overlay hardware.

\subsection{\sysname hardware}
\label{subsec:hardware}
\begin{figure}[!t]
    \centering
    \includegraphics[width=0.9\linewidth]{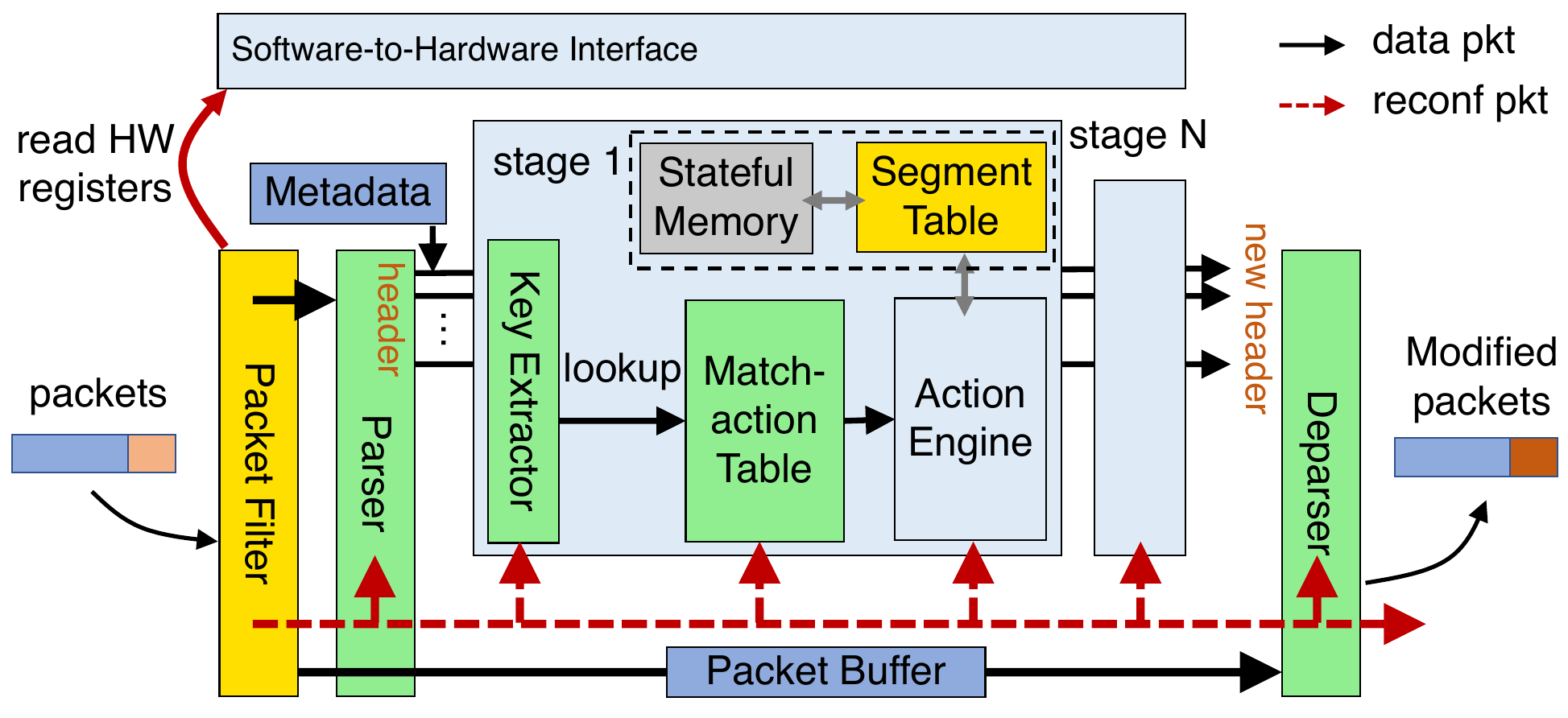}
    \caption{\sysname hardware and software-to-hardware interface. \sysname builds on a RMT~\cite{rmt} pipeline, by adding {\color{goldenyellow}Yellow} components and modifying {\color{applegreen}Green} ones.}
    \label{fig:dataplane}
    \vspace{-.1in}
\end{figure}

The \sysname hardware design (Figure~\ref{fig:dataplane}) builds on RMT by adding hardware primitives for isolation into the RMT pipeline. Because these isolation primitives are added pervasively throughout the pipeline, we first describe the overall \sysname hardware design including both RMT and the new isolation primitives. We then summarize the new isolation primitives added by \sysname.

\sysname expects that a data packet's header carries information identifying what module should process the packet. Currently in our prototype, this is the VLAN ID (VID) header, which we assume is set by the vSwitch~\cite{vl2}, but other fields, e.g., VxLAN ID, can be used instead. Packets entering \sysname are first handled by a packet filter that discards packets without a VLAN ID.\footnote{The filter can be configured to send control packets without VLAN tags, e.g., BFD packets~\cite{bfd}, to the control plane or system-level module (\S\ref{subsec:sys_module}).} Next, a parser extracts the VLAN ID from the packet and applies module-specific parsing to extract module-specific headers from the TCP/UDP payload. The parser then pushes these parsed packet headers into packet header vector (PHV) containers that travel through the pipeline of match-action stages.

Each stage forms keys out of headers, looks up the keys in a match-action table, and performs actions. At the start of each stage, a key extractor in the stage forms a key by combining together the headers in a module-specific manner. The keys are then concatenated with the module ID and looked up in a match-action table, whose space is partitioned across different modules. If the key matches against a match-action pair in the table, the lookup result is used to index an action table.

Similar to the match-action table, the action table is also partitioned across modules. Each action in the table identifies opcodes, operands, and immediate constants for a very-large instruction word (VLIW), controlling many parallel arithmetic and logic units (ALUs). The VLIW instruction consumes the current PHV to produce a new PHV as input for the next stage. The table's action can modify persistent pipeline state, stored in stateful memory. Stateful memory is indexed by a physical address that is computed from a local address, obtained from a module's packets. This computation is done by a segment table, which stores the offset and range of each module's slice of stateful memory. We now detail the main components of our design.

\begin{figure}[!t]
    \centering
    \includegraphics[width=0.9\linewidth]{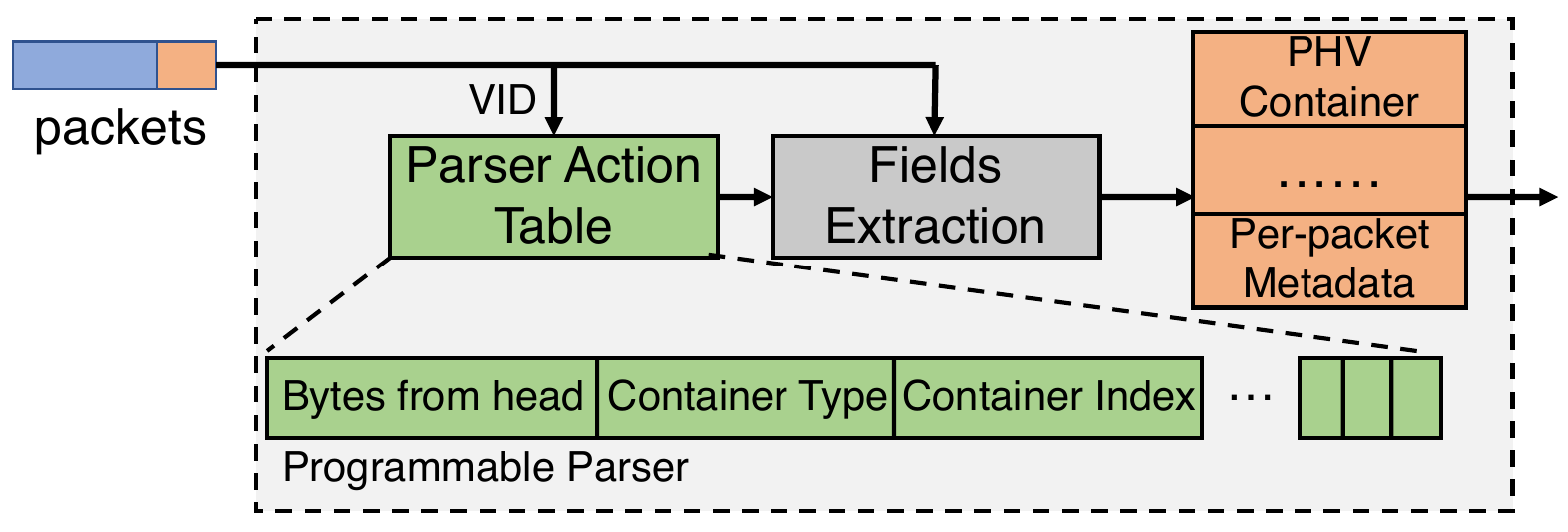}
    \caption{\sysname programmable parser.}
    \label{fig:parser}
\end{figure}

\Para{Parser.} The \sysname parser is driven by a table lookup process similar to the RMT parser~\cite{rmt, gibb_parsing}. Specifically, whenever a new packet comes in, the module ID is extracted from its VLAN ID prior to parsing the rest of the packet. This module ID is then used as an index into the table that determines how to parse the rest of the packet (Figure~\ref{fig:parser}). Each table entry corresponds to multiple parsing actions for a module---one action per extracted PHV container. Each parsing action specifies (1) \textit{bytes from head}, indicating where in the packet the parser should extract a particular header, (2) \textit{container type} (e.g., 4-byte container, etc.), indicating how many bytes we should extract; (3) \textit{container index}, indicating where in the PHV we should put the extracted header into. The parser also sets aside space in the PHV for metadata that is automatically created by the pipeline (e.g., time of enqueue into switch output queues and queueing delay after dequeue) and for temporary packet headers used for computation.

\begin{figure}[!t]
    \centering
    \includegraphics[width=0.8\linewidth]{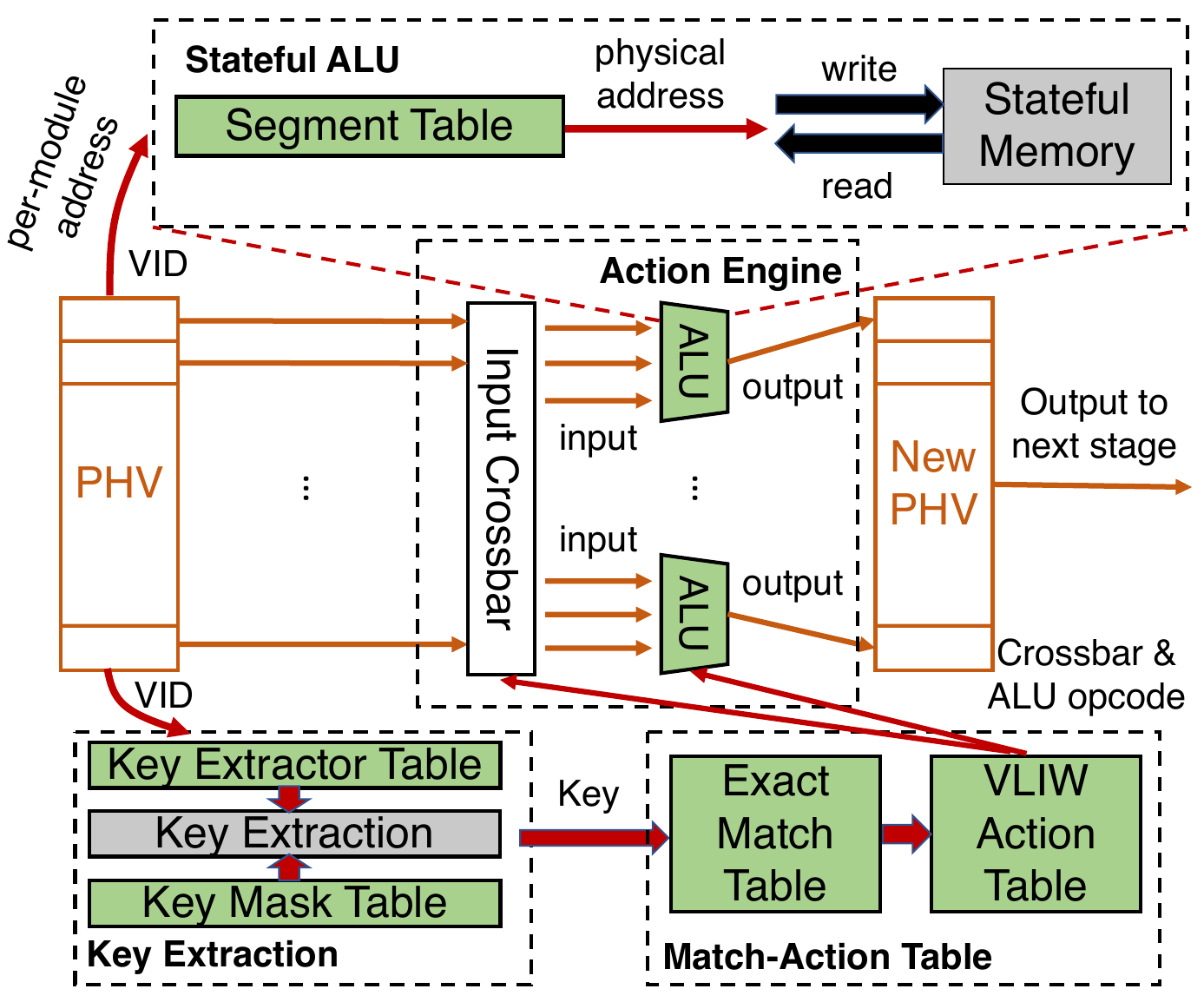}
    \caption{\sysname processing stage.}
    \label{fig:pipeline}
\end{figure}


\Para{Key extractor.} Before a stage performs a lookup on a match-action table, a lookup key must be constructed by extracting and combining together one or more PHV containers. This key extraction process differs between modules in the same stage, and between different stages for the same module. To implement key extraction, just like the parser, we use a key extractor table (Figure~\ref{fig:pipeline}) that is indexed by a packet's module ID. Each entry in this table specifies which PHV containers to combine together to form the key. These PHV containers are then selected into the key using a multiplexer for each portion of the key. To enable variable-length key matching for different modules, the key extractor also includes a key mask table, which also uses the module ID as an index to determine how many bits to pad in the key to bring it up to a certain fixed key size before lookup.

\Para{Match table.} Each stage looks up the fixed-size key constructed by the key extractor in a match table. Currently, we support only exact-match lookup. The match table is statically partitioned across modules by giving a certain number of entries to each module. To enforce isolation among different modules, the module ID is appended to the key output by the key extractor. This augmented key is what is actually looked up against the entries in the match table; each entry stores both a key and the module ID that the key belongs to. The lookup result is used as index into the VLIW action table to identify a corresponding action to execute.

\Para{Action table and action engine.}
Each VLIW action table entry indicates which fields from the PHV to use as ALU operands (i.e., the configuration of each ALU's operand crossbar) and what opcode should be used for each ALU controlled by the VLIW instruction (i.e., addition, subtraction, etc.). Each ALU outputs a value based on its operands and opcode. There is one ALU per PHV container, removing the need for a crossbar on the output because each ALU's output is directly connected to its corresponding PHV container. After a stage's ALUs have modified its PHV, the modified PHV is passed to the next stage.

\Para{Stateful memory.}
\sysname's action engines can also modify persistent pipeline state on every packet. Each module is assigned its own address space, and the available stateful memory in \sysname is partitioned across modules. When a module accesses its slice of stateful memory, it supplies a per-module address that is translated into a physical address by a segment table before accessing the stateful memory. To perform this translation, \sysname stores per-module configuration (i.e., base address and range) in a segment table, which can be indexed by the packet's module ID. \sysname borrows this idea of a segment table from NetVRM's~\cite{netvrm, multitenant_hotcloud} page table, but implements it in hardware instead of programming it in P4 atop Tofino's stateful memory like NetVRM does. This allows \sysname to avoid using scarce Tofino stateful memory to emulate a segment table. Also, by adding segment table hardware to each stage, \sysname avoids sacrificing the first stage of stateful memory for a segment table, instead reclaiming it for useful packet processing. This is unlike NetVRM, which can share stateful memory across modules only from the second stage because the first stage is used for the page table.

\Para{Deparser.}
The deparser performs the inverse operation of the parser. It takes PHV containers and writes them back into the appropriate byte offset in the packet header, merges the packet header with the corresponding payload in the packet buffer, and transmits the merged packet out of the pipeline. The format of the deparser table is identical to the parser table and is similarly indexed by a module ID.

\Para{Secure reconfiguration.}
Our threat model assumes that the \sysname hardware and software are trusted, but that data packets that enter the \sysname pipeline are untrusted. Data packets are untrusted because for a switch, they can come from physical machines outside the switch's control and, for a NIC, they can come from tenant VMs sharing the NIC. Hence, the pipeline should be reconfigured only by \sysname software, not data packets.

This is a security concern faced by existing RMT pipelines as well, even without isolation support. Commercial programmable switches solve this problem by using a separate daisy chain~\cite{daisy_chain} to configure pipeline stages. This chain carries configuration commands that are picked up by the intended pipeline stage as the command passes that stage. The chain is only accessible over PCIe, which is connected to the control-plane CPU, but not by Ethernet ports, which carry outside data packets. Hence, the only way to {\em write} new configurations into the pipeline is through PCIe. The packet-processing pipeline is restricted to just {\em reading} configurations and using them to implement packet processing. Thus, the daisy chain provides secure reconfiguration by physically separating reconfiguration and packet processing.

\sysname uses a similar approach by employing a daisy chain for reconfiguration when a module is updated. A special \textit{reconfiguration packet} carries configuration commands for the pipeline's resources (e.g., parser). Our implementation of this daisy chain varies depending on the platform. For our NetFPGA prototype, this daisy chain is connected solely to the switch CPU via PCIe, similar to current switches. For our Corundum NIC prototype, we connect the daisy chain directly to PCIe and use a {\em packet filter} before our parser to filter out reconfiguration packets from untrusted data packets by ensuring that reconfiguration packets have a specific UDP destination port. An ideal solution would be to use a physically separate interface, e.g., USB or JTAG, for reconfiguring the \sysname pipeline on Corundum, but we found it challenging to implement such a physically separate reconfiguration interface on Corundum. In Appendix~\ref{app:daisy}, we show how a daisy chain permits more rapid reconfiguration than an alternative approach of using the AXI-L protocol on an FPGA.

\Para{Summary of \sysname's new primitives.} The hardware primitives introduced by \sysname on top of an RMT pipeline (Figure~\ref{fig:dataplane}) are the configuration tables for the parser, deparser, key extractor, key mask units and segment table. These tables provide an overlay feature to share the same unit across multiple modules. Specifically, for each unit, \sysname provides a table with a configuration entry per module, rather than one configuration for the whole unit. In addition, \sysname introduces the packet filter to ensure secure reconfiguration. \sysname also modifies match tables, by appending the module ID to the match key and the match-action entries. Finally, \sysname partitions match-action tables and stateful memory across all modules. These primitives ensure that updating one module only affects a single entry (for \sysname resources that use overlays) and only affects a subset of memory (for \sysname resources that use space partitioning), thus allowing us to update one module without disrupting others (\S\ref{ssec:requirements}). 

\Para{ASIC feasibility of \sysname's primitives.}
\label{ssec:asic}
\sysname's parser, deparser, key extractor, key mask, and segment tables are small and simple arrays indexed by the module identifier. They can be readily realized in SRAM that can support a memory read every clock cycle. The packet filter is a simple combinational circuit that checks if the incoming packet is destined to a specific UDP destination port. Extending the match-action tables in each stage to append a module ID to every entry amounts to modestly increasing the key width in the table. While these new primitives add some additional latency relative to RMT, e.g., to go through the packet filter or reading out the per-module parser configuration, the pipelined nature of RMT means that this additional latency does not impact packet-forwarding rate. 
The ASIC area overhead increases as we increase the number of simultaneous programming modules that need to be supported; we quantify it in \S\ref{ssec:perf}.

\subsection{Improving \sysname's throughput}
\label{ssec:optimization}

\begin{figure}[!t]
    \centering
    \includegraphics[width=\linewidth]{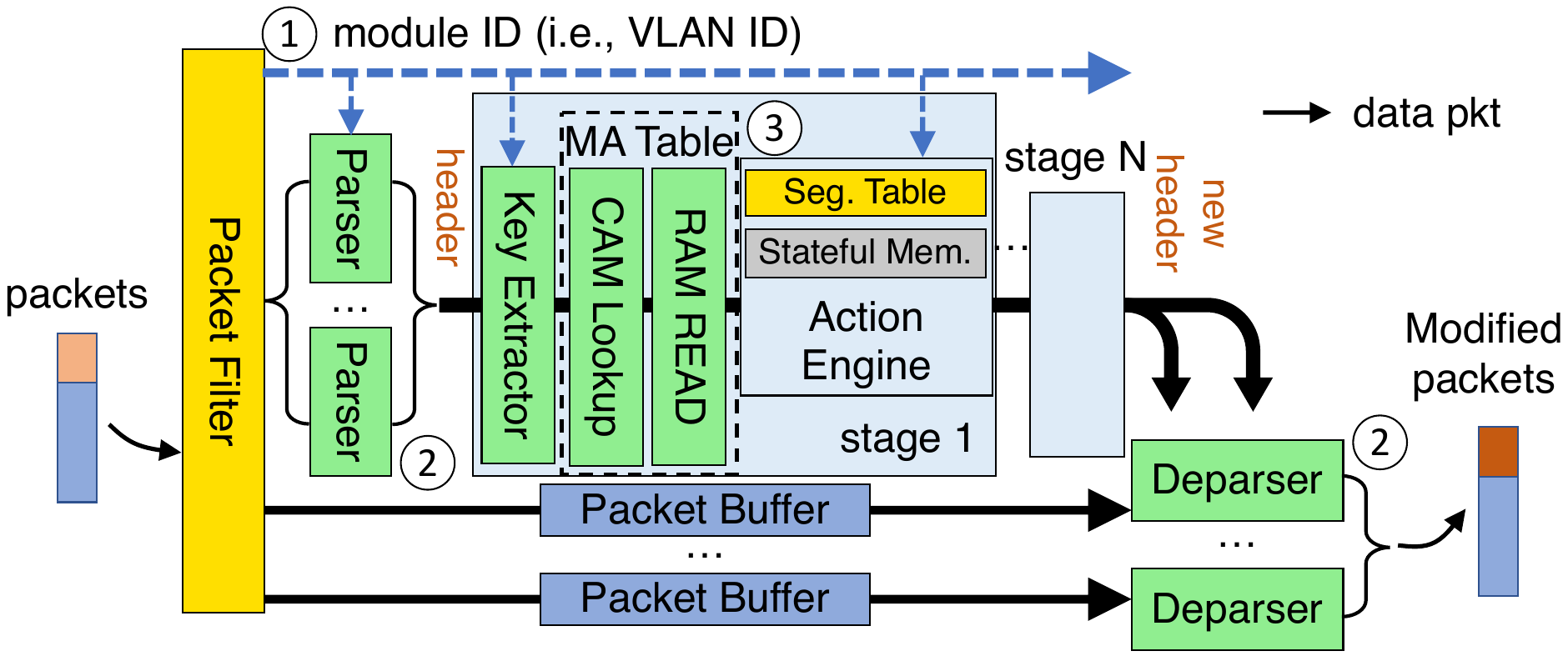}
    \caption{Three optimization techniques applied in \sysname. Numbered circles refer to specific techniques in \S\ref{ssec:optimization}.}
    \label{fig:optimization}
\end{figure}

As shown in Figure~\ref{fig:optimization}, we apply 3 main techniques to optimize the forwarding performance of \sysname: (1) masking RAM read latency, (2) using multiple parsers and deparsers, and (3) increasing pipeline depth. We demonstrate the effect these techniques have on \sysname's throughput in \S\ref{ssec:perf}.

\Para{\ding{172} Masking RAM read latency.} The design described in \S\ref{subsec:hardware} attaches the module ID to the PHV that is sent from one element (e.g., parser, key extractor) to the next. In this design, we read the module's configuration from SRAM after the PHV arrives, thus incurring a few additional clock cycles of latency.
To optimize this, we mask SRAM access latency by splitting the module ID from the PHV and sending the module ID to the next element {\em ahead of time}. The PHV follows the module ID, and thus the module configuration at a stage can be read concurrently with the PHV being transmitted to that stage.



\Para{\ding{173} Multiple parsers and deparsers.} In \S\ref{subsec:hardware}'s design, there is one parser, deparser, and packet buffer. The parser extracts and parses the header and puts the full packet in the packet buffer. Then the deparser takes the modified headers from the last stage, uses them to overwrite the relevant portions of the full packet in the packet buffer, and sends out the packet.

Our optimized design uses multiple parallel parsers, deparsers, and packet buffers to improve throughput. Deparsing is the most expensive operation as any position within the PHV container might be modified, and thus any part of the packet header (128 bytes in our implementation) might need to be updated. Furthermore, deparsing has to process both the packet header and the payload. Therefore, we use 4 parallel deparsers and 2 parsers. We also associate  a separate packet buffer with each deparser.

On ingress, the packet filter tags each packet with a packet buffer number (0--3) in round robin order. It also round robins incoming packets to the 2 parsers. The last pipeline stage uses the packet buffer tag to determine which packet buffer's packet the last stage's modified PHV should be combined with. Each packet buffer's deparser combines the earliest packet from the packet buffer along with the last stage's most recently modified PHV for that buffer.

\Para{\ding{174} Deep pipelining.} With careful digital design, in \sysname's implementation, we can pipeline each element (e.g., match-action table) into several sub elements to improve throughput. For example in Figure~\ref{fig:optimization}, we divide the match-action table into CAM-lookup and action-RAM-read sub elements. In this specific example, this allows us to process a PHV every 2 clock cycles at each sub-element rather than every 4 clock cycles at the whole match-action table.





\begin{figure}[!t]
    \centering
    \includegraphics[width=0.95\linewidth]{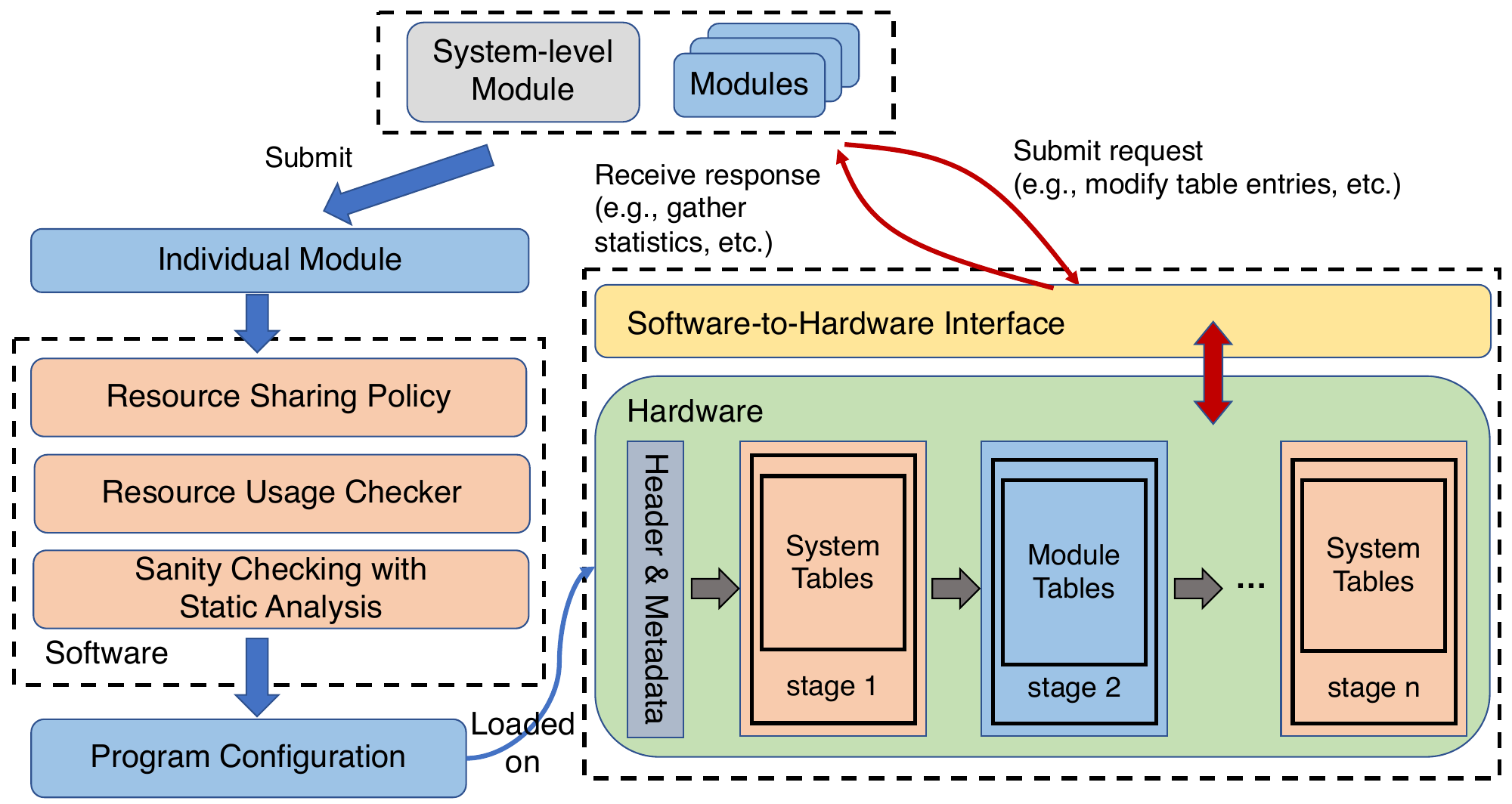}
    \caption{\sysname software and system-level module.}
    \label{fig:sys_overview}
\end{figure}

\subsection{The \sysname system-level module}
\label{subsec:sys_module}

To hide information about the underlying physical infrastructure (e.g., topology) from tenant modules in a virtualized environment, modules in \sysname can use virtual IP addresses to operate in a shared environment~\cite{vl2}. Here, virtual IP addresses are local and scoped to modules belonging to a particular tenant, regardless of which physical device these modules are on. To support virtual IPs and provide basic services to other modules, \sysname contains a system-level module written in P4-16 that provides common OS-like functionality, e.g., converting virtual IPs to physical IPs, multicast,  and looking up physical IPs to find output ports. The system-level module has 3 benefits: (1) it avoids duplication among different modules re-implementing common functions, improving the resource efficiency of the pipeline, (2) it hides underlying physical details (e.g., topology) from each module so that one tenant's modules on different network devices can form a virtual network~\cite{vl2}, and (3)  it provides common and useful real-time statistics (e.g., link utilization, queue length, etc.) that can inform packet processing within modules.

Figure~\ref{fig:sys_overview} shows how the system-level module is laid out relative to the other modules. Packets entering the \sysname pipeline are first processed by the system-level module before being handed off to their respective module for module-specific processing. After module-specific processing, these packets enter the system module for a second time before exiting the pipeline. The first time they enter the system-level module, packets can read and update system-level state (e.g., link utilization, packet counters, queue measurements), whereas the second time they enter the system-level module, module-specific packet header fields (e.g., virtual IP address) can be read by the system-level module to determine device-specific information (e.g., output port). In both halves, there is a narrow interface by which modules communicate with the system-level module. This split structure of the system-level module arises directly from the feed-forward nature of the RMT pipeline, where packets typically only flow forward, but not backward. Hence, packets pick up information from the system-level module in the first stage and pass information to the system-level module in the last stage. The non-system modules are sandwiched in between these two halves.


\subsection{\sysname software}
\label{subsec:software}

\Para{The software-hardware interface.} The \sysname software-to-hardware interface works similar to P4Runtime~\cite{p4runtime} to support interactions (e.g., modifying match-action entries, fetching hardware statistics, etc.) between the \sysname software and the \sysname hardware. However, in addition to P4Runtime's functions, \sysname's software-hardware interface can also be used to reconfigure different hardware resources (Appendix~\ref{app:considered_hw_resources}) in \sysname to reprogram them when a module is added or updated. This allows us to dynamically reconfigure portions of \sysname as module logic changes.

\Para{The \sysname resource checker.} The \sysname resource checker ensures that each module's resource allocation complies with an operator specified resource sharing policy (e.g., dominant resource sharing (DRF)~\cite{drf}, or a utility-based~\cite{nsdi2022hogan} policy). In our current design we check allocations statically because reassigning resources from one module to another disrupts processing for both modules. Instead we rely on admission control and do not load a module whose resource requirements cannot be met. We leave the question of what is an appropriate resource allocation policy to future work.

\Para{The \sysname static checker.} To ensure isolation, \sysname's static checker analyzes 3 properties of the module's P4 source code. First, it checks that modules do not modify hardware-related statistics (e.g., link utilization) provided by the system-level module to all modules. Second, modules can not modify their VID. This is because a module can be spread across multiple programmable devices~\cite{netchain, lyra}, and changes to VIDs by module $A$ on a device can unintentionally affect a module $B$ on a downstream device, where $B$'s real VID happens to be the same as $A$'s modified VID. Third, modules must not recirculate packets and their routing tables should be loop-free.\footnote{We check loop freedom in the control plane.} This is because all modules share the same ingress pipeline bandwidth. Recirculating packets or looping them back through multiple devices will degrade the ability of other modules to process packets.

\Para{The \sysname compiler.} Packet-processing pipelines (e.g., RMT~\cite{rmt}) are structured as feed-forward pipelines of programmable units, each of which has limited processing capabilities. This design ensures the {\em all-or-nothing} property: once a module has been compiled and loaded it can run at up to line rate, while modules that can not run at line rate cannot be compiled. \sysname's compiler follows the same design, and only admits modules that meet line-rate requirements.

The compiler reuses the frontend and midend of the open-source P4-16 reference compiler~\cite{p4c} and creates a new backend similar to BMv2~\cite{bmv2_model}. This backend has a parser, a single processing pipeline, and a deparser. The compiler takes a module's P4-16 program as input and conducts all the resource usage and static checks described above. Then, for the parser and deparser, it transforms the parser defined in the module to configuration entries for the parser and deparser tables. For the packet-processing pipeline, which consists of match-action tables, it transforms the key in a table to a configuration in the key extractor table, and actions to VLIW action table entries according to the opcodes. The compiler also performs dependency checking~\cite{rmt, lavanya_compiler} to guarantee that all ALU actions and key matches are placed in the proper stage, respecting table dependencies.

The \sysname compiler can be extended to support the same packet flowing through different P4 modules belonging to one tenant. The compiler can take multiple P4 modules as input, assign them the same module ID, and allocate them to non-overlapping pipeline stages---similar to how we lay out user and system modules in different stages as in Figure~\ref{fig:sys_overview}.

\subsection{Limitations}

As a research prototype, \sysname has several limitations. First, while we have developed {\em mechanisms} to support isolation across multiple modules, we have not yet designed {\em policies} that decide how much of each resource a module should be given~\cite{switchhire}. Second, our FPGA implementation of RMT lacks many features present in a commercial RMT implementation such as the Barefoot Tofino switch~\cite{tofino}. Third, our compiler currently does not perform any compiler optimizations for code generation~\cite{chipmunk} or memory allocation~\cite{lavanya_compiler, lyra}. Fourth, \sysname proposes isolation mechanisms for the packet-processing pipeline, but does not deal with isolating traffic from different modules competing for output link bandwidth, which is a orthogonal traffic management problem. Proposals like PIFO~\cite{pifo} can be used here, by assigning PIFO ranks to different modules to realize a desired inter-module bandwidth-sharing policy.
\section{Implementation}
\label{sec:implementation}

\begin{figure}[!t]
    \centering
    \includegraphics[width=0.8\linewidth]{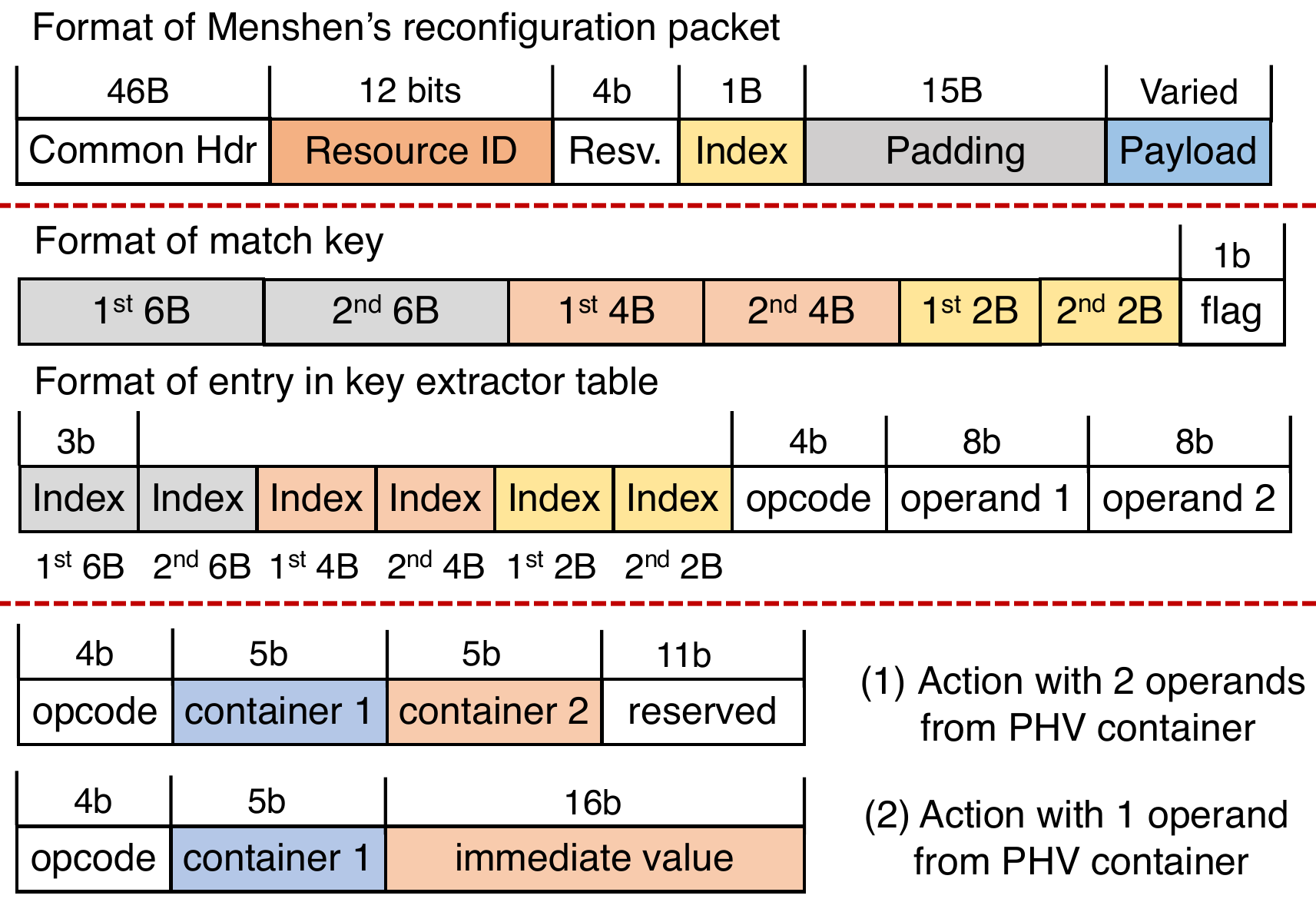}
    \caption{Formats of \sysname's packets and tables.}
    \label{fig:impl_format}
\end{figure}

\subsection{\sysname hardware}
\label{subsec:impl_hardware}


To implement \sysname, we first built a baseline RMT implementation for an FPGA. \sysname includes (1) a packet filter to filter out reconfiguration packets from data packets using a specific predefined UDP destination port (i.e., 0xf1f2), (2) a programmable parser, (3) a programmable RMT pipeline with 5 programmable processing stages, (4) a deparser, and (5) a separate daisy-chain pipeline for reconfiguration. It also includes \sysname's primitives for isolation. We have integrated it into both the Corundum NIC \cite{corundum} and the NetFPGA reference switch \cite{netfpga}. The \sysname code base together with the optimizations (\S\ref{ssec:optimization}) consists of 9975 lines of Verilog. Of this, 3098 and 3226 lines are for handling data bus widths of 512 bits (Corundum) and 256 bits (NetFPGA) respectively. 3651 lines are for the common blocks, e.g., key extractor, etc. Below, we describe our hardware implementation in more detail. Figure~\ref{fig:impl_format} shows the formats of \sysname's packets and tables.

\Para{PHV format.} \sysname's PHV has 3 types of containers of different sizes, namely 2-byte, 4-byte and 6-byte containers. Each type has 8 containers. Also, we allocate and append an additional 32 bytes to store platform-specific metadata (e.g., an indication to drop the packet, destination port, etc.), which results in a PHV length of 128 bytes in total. Thus, we have a total of $3*8+1=25$ PHV containers. To prevent any possibility of PHV contents leaking from one module to another, the PHV is zeroed out for each incoming packet.

\Para{Reconfiguration packet format.} Figure~\ref{fig:impl_format} shows the format of \sysname reconfiguration packets. The reconfiguration packet is a UDP packet with the standard UDP, Ethernet, VLAN, and IP headers. Within the UDP payload, a 12-bit resource ID indicates which hardware resource within which stage should be updated (e.g., key extractor table in stage 3). To reconfigure the resource, the table storing the configuration for this resource must be updated by writing the entry stored within the reconfiguration packet's payload at the location specified by the 1-byte index field in the reconfiguration packet header. The UDP destination port field determines whether the reconfiguration packet is valid or not.              

\Para{Packet filter.} The packet filter has 2 registers that can be accessed by the \sysname software via Xilinx's AXI-Lite protocol~\cite{axil}: (1) a 4-byte reconfiguration packet counter, which monitors how many reconfiguration packets have passed through the daisy chain; (2) a 32-bit bitmap, which indicates which module is currently being updated (e.g., bit 1 stands for module 1, bit 2 for module 2, etc.). During reconfiguration of a module, via the software-to-hardware interface, the \sysname software reads the reconfiguration packet counter. It then writes the bitmap to reflect the module ID $M$ of the module currently being updated. The bitmap is then consulted on every packet to drop data packets from $M$ until reconfiguration completes, so that $M$'s ``in-flight'' packets aren't incorrectly processed by partial configurations.

Then, the \sysname software sends all reconfiguration packets embedded with the predefined UDP destination port to the daisy chain. Finally, it polls the reconfiguration packet counter to check if reconfiguration is over and then zeroes the bitmap so that $M$'s packets are no longer dropped. Reconfiguration packets maybe dropped before they reach the RMT pipeline. This can be detected by polling the reconfiguration packet counter to see if it has been correctly incremented or not. If it hasn't been incremented correctly, then the entire reconfiguration process restarts with $M$'s packets being dropped until reconfiguration is successful.

\begin{table}[!t]
\scriptsize
    \begin{tabular}{ll}
        \toprule
        \textbf{Operation} & \textbf{Description} \\
        \midrule 
        \texttt{add/sub} & Add/subtract between containers \\
        \texttt{addi/subi} & Add/subtract an immediate to/from container \\
        \texttt{set} & Set a container to an immediate value \\
        \texttt{load} & Load a value from stateful memory \\
        \texttt{store} & Store a value to stateful memory \\
        \texttt{loadd} & Load value from stateful memory, add 1, and store back \\
        \texttt{port} & Set destination port \\
        \texttt{discard} & Discard packet \\
        \bottomrule
    \end{tabular}
\caption{Supported operations in \sysname's ALU.}
\label{tab:supported_ops}
\end{table}

\Para{Programmable parser/deparser.} We currently support per-module packet header parsing in the first 128 bytes of the packet. These 128 bytes also include the headers common to all modules (e.g., Ethernet, VLAN, IP, and UDP). We design the parser action for each parsed PHV container as a 16-bit action. The first 3 bits are reserved. The next 7 bits indicate the starting extraction position in bytes from byte 0. These 7 bits can cover the whole 128-byte length. Then, the next 2 bits and 3 bits indicate the container type (2, 4, or 6 byte) and number (0--7) respectively. The last bit is the validity bit. For each module, we allocate 10 such parser actions (i.e., to parse out at most 10 containers), resulting in a 160-bit-wide entry for the parser action table.

We note that we only parse out fields of a packet into PHV containers, if those fields are actually used as part of either keys or actions in match-action tables. Before packets are sent out, the deparser pulls out the full packet (including the payload) from the packet buffer and only updates the portions of the packet that were actually modified by table actions. This approach allows us to reduce the number of PHV containers to 25 because packet fields that are never modified or looked up by the \sysname pipeline need not travel along with the PHV.

\Para{Key extractor.} The key for lookup in the match-action table is formed by concatenating together up to 2 PHV containers each of the 2-byte, 4-byte, and 6-byte container types. Hence the key can be up to 24 bytes and 6 containers long. Since there are 8 containers per type, the key extraction table entry for each module in each stage uses $log_{2}(8)*6=18$ bits to determine which container to use for the 6 key locations.
Additionally, the key extractor is also used to support conditional execution of actions based on the truth value of a predicate of the form $A~OP~B$, where $A$ and $B$ are packet fields and $OP$ is a comparison operator. For this purpose, each key extractor table entry also specifies the 2 operands for the comparison operation and the comparison opcode. The opcode is a 4-bit number, while the operands are 8 bits each. The operands can either be an immediate value or refer to one of the PHV containers. The result of the predicate evaluation adds one bit to the original 24 byte key, bringing the total key length to $24*8 + 1 = 193$ bits.
Because not all keys need to be 193 bits long, we use a 193-bit-wide mask table. Each entry in this table denotes the validity of each of the 193 key bits for each module in each stage. This is somewhat wasteful and can be improved by storing validity information within the key extractor table itself.

\Para{Exact match table.} To implement the exact match table, we leverage the Xilinx CAM block~\cite{xilinx_cam}. This CAM matches the key from the key extractor module against the entries within the CAM. As discussed in~\cref{subsec:hardware}, to ensure isolation between different modules, we append the module ID (i.e., VLAN ID) to each entry, which means that the CAM has a width of $193+12=205$ bits. The lookup result from the CAM is used to index the VLIW action table. The action is designed in a 25-bit format per ALU/container (Figure~\ref{fig:impl_format}). As we have $24+1=25$ PHV containers, the width of the VLIW action table is $25*25=625$ bits. The Xilinx CAM block simplifies implementation of an exact-match table and can also easily support ternary matches if needed (Appendix~\ref{app:tcam}). 

\Para{Action engine.} The crossbar and ALUs in the action engine use the VLIW actions to generate inputs for each ALU and carry out  per-ALU operations. ALUs support simple arithmetic, stateful memory operations (e.g., loads and stores), and platform-specific operations (e.g., discard packets) (Table~\ref{tab:supported_ops}). The formats of these actions are shown in Figure~\ref{fig:impl_format}. Additionally, in stateful ALU processing, each entry in the segment table is a 2-byte number, where the first byte and second byte indicate memory offset and range, respectively.

\Para{\sysname primitives.} \sysname's isolation primitives (e.g., key-extractor and segment tables) are simple arrays implemented using the Xilinx Block RAM~\cite{xilinx_ram} feature.


\subsection{\sysname Software}
The \sysname compiler reuses the open-source P4-16 reference compiler~\cite{p4c} and implements a new backend extension in 3773 lines of C++. It takes the module written in P4-16 together with resource allocation as the inputs, and generates per-module configurations for \sysname hardware.  Specifically, it (1) conducts resource usage checking to ensure every program's resource usage is below its allocated amount; (2) places the system-level module's (120 lines of P4-16) configurations in the first and last stages in the \sysname pipeline; and (3) allocates PHV containers to the fields shared between the system-level and other modules so that the other modules can be sandwiched between the two halves of the system-level module (\cref{subsec:software}). The \sysname software-to-hardware interface is written in Python. It configures \sysname hardware by converting program configurations to reconfiguration packets.

\subsection{Corundum and NetFPGA integrations}
\label{subsec:impl_setup}

We have integrated \sysname into 2 FPGA platforms: one for the NetFPGA platform that captures the hardware architecture of a switch~\cite{netfpga}, and another for the Corundum platform that captures the hardware architecture of a NIC~\cite{corundum}. \sysname's integration on Corundum~\cite{corundum} is based on a 512-bit AXI-S \cite{axi4_stream} data width and runs at 250 \si{MHz}. Although \sysname's pipeline can be integrated into both the sending and receiving path, in our current implementation, we have integrated \sysname into only Corundum's sending path, i.e., PCIe input to Ethernet output. \sysname on NetFPGA~\cite{netfpga} uses a 256-bit AXI-S \cite{axi4_stream} data width and runs at 156.25 \si{MHz}. 

On the Corundum NIC platform, we insert a 1-bit discard flag, while on the NetFPGA switch platform, we insert a 1-bit discard flag and 128-bit platform-specific metadata, i.e., source port, destination port and packet length, into the PHV's metadata field. A 4-bit one-hot encoded tag indicates the packet buffer (\S\ref{ssec:optimization}). The table depth in \sysname's parser, key extractor, key mask, page, and deparser tables affects the maximum number of modules we can support and is currently 32. The depth of CAM and VLIW action table directly influences the amount of match-action entries and VLIW actions that can be allocated to all modules. Due to the open technical challenge of implementing CAMs on FPGAs efficiently \cite{flowblaze, fpga_tcam_limit}, we set their depth to 16 in each stage. While 16 is a small depth, the depth can be improved by using a hash table, rather than a CAM, for exact matching, e.g., cuckoo hashing~\cite{cuckoo}.

\section{Evaluation}
\label{sec:evaluation}
\begin{table}[!t]
\begin{center}
\scriptsize 
    \begin{tabular}{ll}
        \toprule
        \textbf{Program} & \textbf{Description} \\
        \midrule
        CALC~\cite{p4tutorial} & return value based on parsed opcode and operands \\
        Firewall~\cite{p4tutorial} & stateless firewall that blocks certain traffic \\
        Load Balancing~\cite{p4tutorial} & steer traffic based on 4-tuple header info \\
        QoS~\cite{p4tutorial} & set QoS based on traffic type \\
        Source Routing~\cite{p4tutorial} & route packets based on parsed header info \\
        NetCache~\cite{netcache} & in-network key-value store \\
        NetChain~\cite{netchain} & in-network sequencer \\
        Multicast~\cite{p4tutorial} & multicast based on destination IP address \\ 
        \bottomrule
    \end{tabular}
\caption{Evaluated use cases.}
\label{tab:evaluated_usecases}
\end{center}
\vspace{-0.2in}
\end{table}


In \S\ref{ssec:req_eval}, we show that \sysname can meet our requirements (\S\ref{ssec:requirements}): it can be rapidly reconfigured, is lightweight, provides behavior isolation, and is disruption-free. \sysname achieves performance isolation by (1) assuming packets exceed a minimum size (to guarantee line rate) and (2) forbidding recirculation. If either is violated, hardware rate limiters can be used to limit each module's packet/bit rate. It achieves resource isolation by ensuring that a table entry for a resource (e.g., parser) is allotted to at most one module. In \S\ref{ssec:perf}, we evaluate the current performance of \sysname.

\Para{Experimental setup.}
To demonstrate \sysname's ability to provide multi-module support, we picked 6 tutorial P4 programs~\cite{p4tutorial}, as detailed in Table~\ref{tab:evaluated_usecases}, together with simplified versions of NetCache~\cite{netcache} and NetChain~\cite{netchain}.\footnote{Our versions of NetChain and NetCache do not include some features such as tagging hot keys.} The system-level module provides basic forwarding and routing, with multicast logic integrated in it. \sysname's parameters are detailed in
\cref{sec:implementation} and summarized in Table~\ref{tab:parameters} in the Appendix.

\Para{Testbed.} We evaluate \sysname based on our Corundum and NetFPGA integrations as described in \cref{sec:implementation}. For the switch platform experiments on NetFPGA, we use a single quad-port NetFPGA SUME board~\cite{netfpga-sume}, where two ports are connected to a machine equipped with an Intel Xeon E5645 CPU clocked at 2.40 GHz and a dual-port Intel XXV710 10/25GbE NIC. For the NIC platform experiments on Corundum, we use a single Xilinx Alveo U250 board~\cite{au250}, where one port is with \sysname for the transmitting path and this port is connected to a 100 GbE NIC as the receiving path. Both setups are used to check \sysname's correctness (\S\ref{ssec:req_eval}). 
For NetFPGA performance tests (\S\ref{ssec:perf}), we use the host as a packet generator.
For Corundum performance tests (\S\ref{ssec:perf}), we internally connect its receiving and transmitting path, and use the Spirent tester~\cite{spirent_tester} to generate traffic. We depict our testing setup in Appendix~\ref{app:tb_setup}.


\subsection{Does \sysname meet its requirements?}
\label{ssec:req_eval}

\begin{figure*}[!thbp]
 \begin{minipage}[t]{0.3\linewidth}
  \centering
  \includegraphics[width=0.95\linewidth]{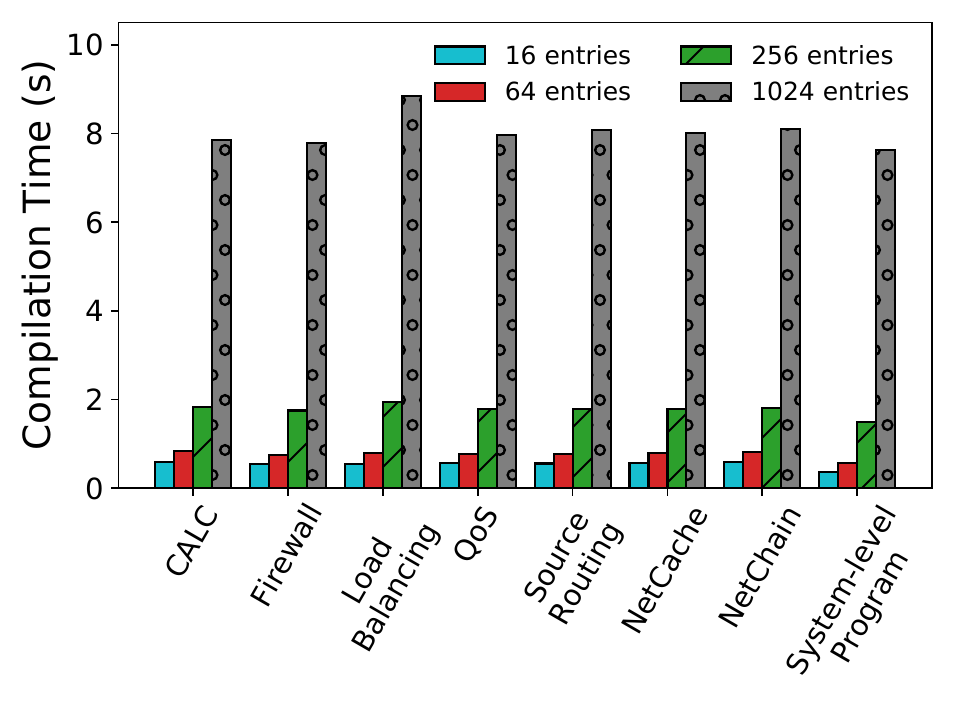}
  \vspace{-0.1in}
  \caption{Compilation time.\label{exp_fig:compilation_time}}
  \vspace{-0.15in}
 \end{minipage}
 \hfill
 \begin{minipage}[t]{0.3\linewidth}
  \centering
  \includegraphics[width=0.95\linewidth]{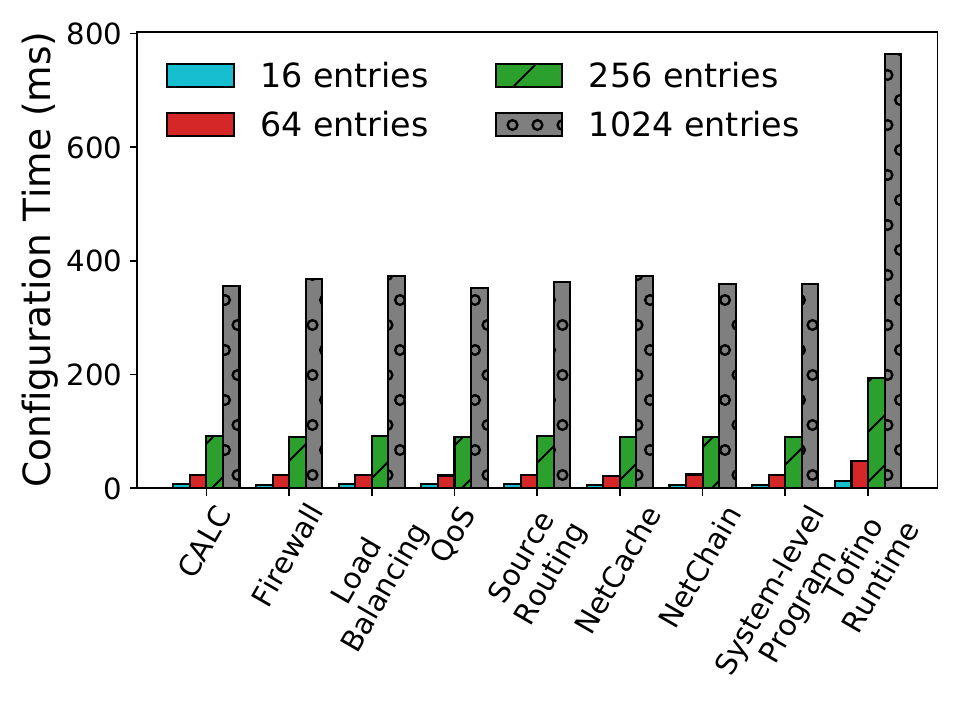}
  \vspace{-0.1in}
  \caption{Configuration time.\label{exp_fig:reconf_time}}
  \vspace{-0.15in}
 \end{minipage}
 \hfill
 \begin{minipage}[t]{0.36\linewidth}
  \centering
  \includegraphics[width=0.85\linewidth]{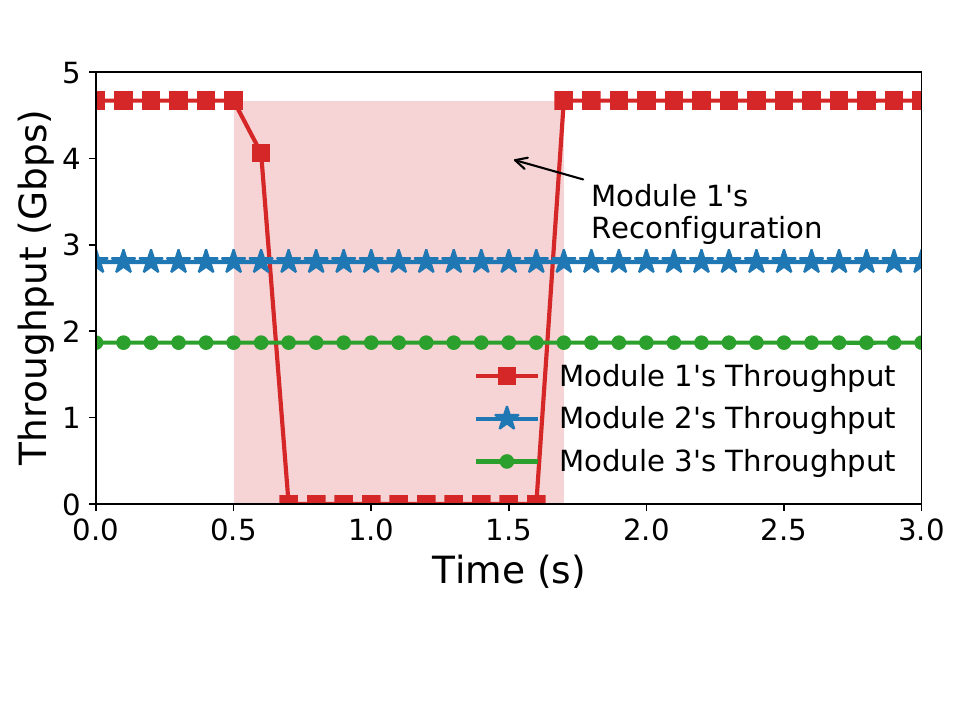}
  \vspace{-0.1in}
  \caption{Throughput during reconfiguration.\label{exp_fig:isolation}}
  \vspace{-0.15in}
 \end{minipage}
 \hfill
\end{figure*}

\Para{\sysname can be rapidly reconfigured.} Reconfiguration time includes both the software's compilation time (Figure~\ref{exp_fig:compilation_time}) and the hardware's configuration time (Figure~\ref{exp_fig:reconf_time}); we evaluate each separately. When a module is compiled, the compiler needs to generate both configuration bits for various hardware resources as well as match-action entries for the tables the module looks up. These match-action entries can and will be overwritten by the control plane, but we need to start out with a new set of match-action entries for a module to ensure no information leaks from a previous module.

Hence, every time a module is compiled, the compiler also generates match-action entries. Within an exact match table, these entries must be different from each other to prevent multiple lookup results. As a result, \sysname's compilation time increases with the number of match-action entries in the module (Figure~\ref{exp_fig:compilation_time}). To contextualize this, \sysname's compile times (few seconds) compare favorably to compile times for Tofino ($\sim$10 seconds for our use cases) and FPGA synthesis times (10s of minutes). We note that this is an imperfect comparison: our compiler performs fewer optimizations than either the Tofino or FPGA compilers and our targets are simpler. That said, compilation can happen offline, and hence it is not as time-sensitive compared to run-time reconfiguration.

To measure time taken for \sysname's configuration post compilation, we vary the number of entries the \sysname software has to write into the pipeline.\footnote{Since the \sysname hardware can't currently support so many entries (\cref{subsec:impl_setup}), we overwrite previously written entries to measure configuration time.} Also, as a comparison, we evaluate the cost of the Tofino run-time APIs from Tofino SDE 9.0.0 to insert match-action table entries for the CALC program. From Figure~\ref{exp_fig:reconf_time}, we observe that the time spent in configuration of the hardware via \sysname's software-to-hardware interface is similar to Tofino's run-time APIs.

\Para{\sysname can reconfigure without disruption.} To show \sysname can support disruption-free reconfiguration, we launch three CALC programs with fixed input packet rate, i.e., 5:3:2 ratio on a single link for module 1, 2 and 3, respectively. We use netmap-based tcprelay to generate total traffic of 9.3 \si{Gbit/s} on a 10 \si{Gbit/s} link. 0.5 seconds in, we start to reconfigure the first module to see if the packet processing of other modules has stalled or not. In Figure~\ref{exp_fig:isolation} we show the throughput achieved by each of three modules when reconfiguring module 1. We can observe that model 2 and 3 see no impact on their throughput. This demonstrates that \sysname provides performance isolation, and that it is feasible for a tenant to reconfigure their module without impacting other tenants. By contrast, updating a module on Tofino (\S\ref{sec:related}) requires resetting the entire switch pipeline. Even with Tofino's Fast Refresh~\cite{tofino_fast_refresh}, this leads to a 50 \si{\milli\second} disruption of all servers (and their VMs) whose traffic is routed through the switch. This disruption can be significant in public cloud environments, and in many cases renders dynamic reconfiguration infeasible.

\begin{table}[!t]
\begin{center}
\scriptsize 
    \begin{tabular}{lrr}
        \toprule
        \textbf{Hardware Implementation }& \textbf{Slice LUTs} & \textbf{Block RAMs} \\
        \midrule 
        NetFPGA reference switch & 42325 (9.77\%) & 245.5 (16.7\%) \\
        RMT on NetFPGA & 200573 (46.3\%) & 641 (43.6\%) \\
        \textbf{\sysname on NetFPGA} & 200733 (46.34\%) & 641 (43.6\%) \\
        Corundum & 61463 (3.56\%) & 349 (12.98\%) \\
        RMT on Corundum & 235686 (13.63\%) & 316 (11.75\%) \\ 
        \textbf{\sysname on Corundum} & 235903 (13.65\%) & 316 (11.75\%) \\
        \bottomrule
    \end{tabular}
\caption{Resources used by 5-stage \sysname pipeline, on NetFPGA SUME and AU250 boards, compared with reference switch, Corundum NIC, and RMT.}
\label{tab:resource_requirement}
\vspace{-0.25in}
\end{center}
\end{table}

\Para{\sysname is lightweight.} We list \sysname's resource usage of logic and memory (i.e., LUTs and Block RAMs), including absolute numbers and fractions, in Table~\ref{tab:resource_requirement}. For comparison, we also list the resource usage of the NetFPGA reference switch and the Corundum NIC. We believe that the additional hardware footprint of \sysname is acceptable for the programmability and isolation mechanisms it provides relative to the base platforms. The reason that \sysname uses more LUTs than Block RAMs is that \sysname leverages the Shift Register Lookup (SRL)-based implementation of Xilinx's CAM IP \cite{xilinx_cam}. We also compared with an RMT design, where we modified \sysname's hardware to support only one module. Relative to RMT, \sysname incurs an extra 0.65\% (NetFPGA) and 0.15\% (Corundum) in LUTs usage.

\Para{\sysname provides behavior isolation.} Next, we spot check that \sysname can correctly isolate modules, i.e., every running module can concurrently execute its desired functionality. For this, we ran the CALC, Firewall, and NetCache module simultaneously on the \sysname pipeline. We generate data packets of different VIDs, which indicate which of these 3 modules they belong to, and input them to the \sysname FPGA prototype on both platforms. By examining the output packets at the end of \sysname's pipeline, we checked that \sysname had correctly isolated the modules, i.e., each module behaved as it would have had it run by itself. We repeated the same experiment by running the Load Balancing, Source Routing, and NetChain modules simultaneously; we observed correct behavior isolation here too.


\subsection{\sysname Performance}
\label{ssec:perf}

\begin{figure*}[!h]
    \centering
    \begin{subfigure}{0.24\linewidth}
        \includegraphics[width=\textwidth]{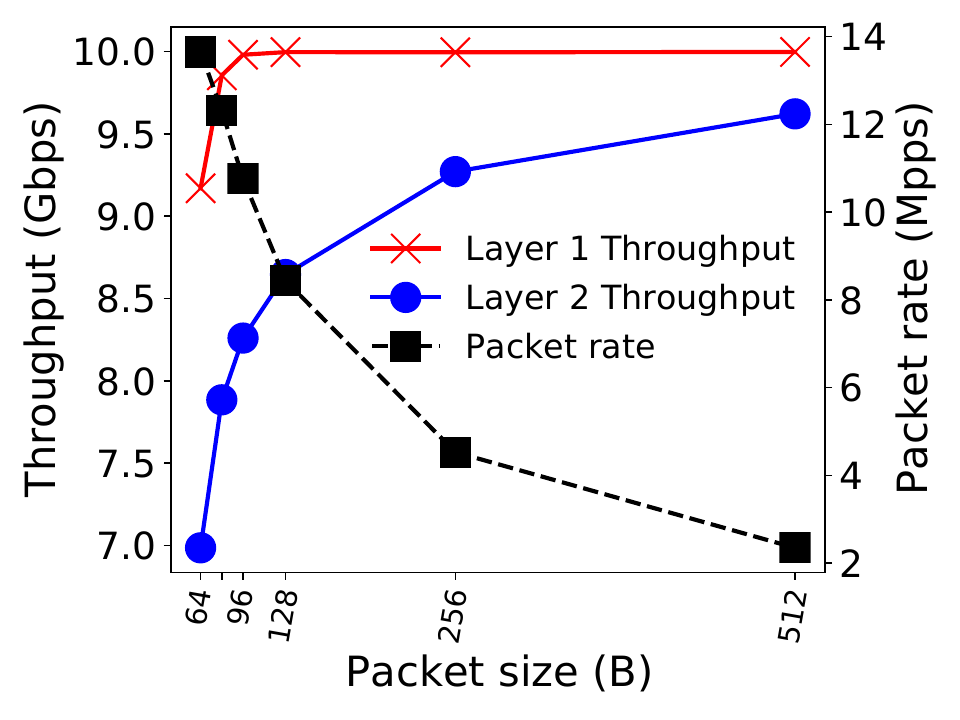}
        \vspace{-0.2in}
        \caption{Optimized NetFPGA.}
        \label{exp_fig:netfpga_perf}
    \end{subfigure} \hfill
    \begin{subfigure}{0.24\linewidth}
        \includegraphics[width=\textwidth]{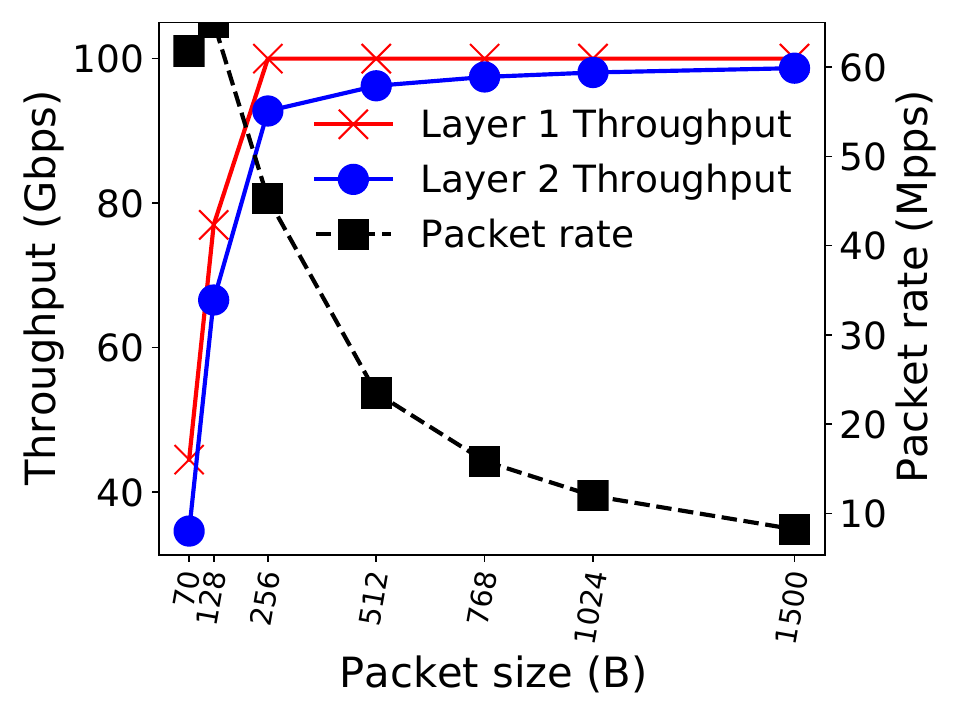}
        \vspace{-0.2in}
        \caption{Optimized Corundum.}
        \label{exp_fig:corundum_perf}
    \end{subfigure} \hfill
    \begin{subfigure}{0.24\linewidth}
        \includegraphics[width=\textwidth]{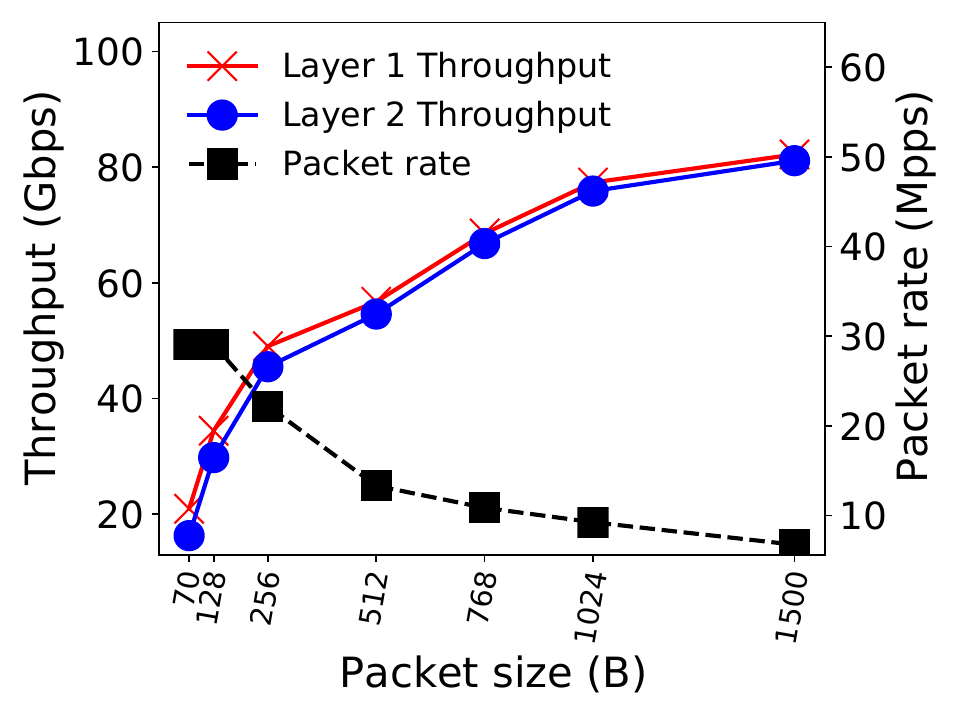}
        \vspace{-0.2in}
        \caption{Unoptimized Corundum.}
        \label{exp_fig:corundum_single_pipe_perf}
    \end{subfigure} \hfill
    \begin{subfigure}{0.24\linewidth}
        \includegraphics[width=\textwidth]{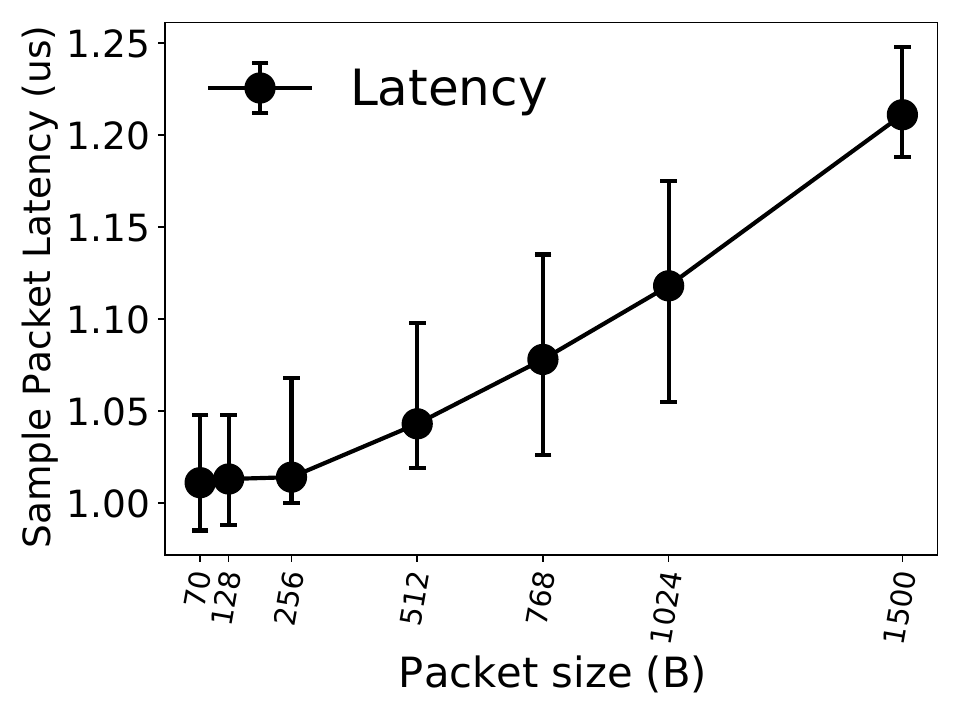}
        \vspace{-0.2in}
        \caption{Optimized Corundum latency.}
        \label{exp_fig:corundum_latency_perf}
    \end{subfigure}
    \caption{Results for performance benchmarks.}
    \label{exp_fig:perf_test}
    \vspace{-0.25in}
\end{figure*}

\Para{How many modules can be packed?} In our current prototype on both Corundum and NetFPGA, we can support at most 32 modules because each isolation primitive (e.g., key extractor table) currently has 32 entries. In practice, the number of modules could be less than 32 if modules need to share a more bottlenecked hardware resource. For instance, if each module wants a match-action entry in every pipeline stage, the maximum number of modules is at most 16 because there are only 16 match-action entries in each stage in our current prototype. However, the numbers above are entirely a function of how much hardware one is willing to pay in exchange for multitenancy support. If we can afford to expend additional resources on an FPGA or extra area on an ASIC, we can correspondingly support a larger number of modules.

\Para{Latency.} In our current implementation, the number of clock cycles needed to process a packet in the pipeline depends on packet size. This is because the number of cycles to process both the header and the payload depend on the header and payload length. For instance, for a minimum packet size of 64 bytes, \sysname's pipeline introduces 79 and 106 cycles of processing for NetFPGA and Corundum, resulting in $79*\frac{1000}{156.25}=505.6$ \si{\nano\second} and $106*\frac{1000}{250}=424$ \si{\nano\second} latency, respectively. For the max. packet size of 1500 bytes, \sysname incurs 146 and 112 cycles for NetFPGA and Corundum, resulting in $150*\frac{1000}{156.25}=960$ \si{\nano\second} and $129*\frac{1000}{250}=516$ \si{\nano\second} latency.

\Para{Throughput.} For NetFPGA, we used MoonGen~\cite{moongen} to generate packets with different sizes. Figure~\ref{exp_fig:netfpga_perf} shows that \sysname achieves a rate of 10 Gbit/s after a packet size of 96 bytes. This is the maximum supported by our MoonGen setup because we have a single 10G NIC. For Corundum, we internally connected Corundum's receiving and transmitting path. Rather than using a host-based packet generator through PCIe, we used Spirent FX3-100GO-T2 tester to test \sysname's throughput. The MTU size is set to 1500 bytes. As shown in Figure~\ref{exp_fig:corundum_perf} and Figure~\ref{exp_fig:corundum_single_pipe_perf}, optimized \sysname on Corundum achieves 100 \si{Gbit/s} at 256 bytes, while unoptimized \sysname can only achieve 80 \si{Gbit/s} at MTU-size packets. 
Also, we sample packets to evaluate the packet latency of optimized \sysname on Corundum with full rate. As depicted in Figure~\ref{exp_fig:corundum_latency_perf}, at full rate, it incurs about 1.2 \si{\micro\second} latency.


\Para{ASIC feasibility.} With the same parameter settings in \S\ref{sec:evaluation}, we use the Synopsys DC synthesis tool~\cite{synopsys_dc} and FreePDK45nm technology library~\cite{freepdk45} to assess the ASIC feasibility of the \sysname pipeline.\footnote{Since we can not have access to source code of Xilinx IPs (e.g., DMA, Ether+PHY, etc.), we solely run synthesis on \sysname's Verilog codebase.} At 1 GHz frequency, when compared with an RMT design, where we modified \sysname to support only one module, \sysname incurs 18.5\%, 7\%, 20.9\% additional chip area for the parser, deparser and one stage, respectively. For a 5-stage pipeline along with the packet filter, parser, deparser and packet buffers, \sysname (10.81 \si{\milli\meter\squared}) incurs 11.4\% additional chip area compared with RMT (9.71 \si{\milli\meter\squared}).

Considering that memory (i.e., lookup tables) and packet processing logic only costs at most 50\% in switch chip area~\cite[page~36]{pkt_processing_chiparea}, \sysname's chip area overhead is moderate ($11.4\%*50\%=5.7\%$), which is conservative since the number of entries in our match-action table is only 16 (\S\ref{subsec:impl_hardware}). With much larger number of entries in lookup tables---which is the common block between \sysname and RMT---\sysname's additional chip area will be negligible.

\section{Related work}
\label{sec:related}


\Para{Multi-core architecture solutions.} To support isolation on programmable network devices based on multicores~\cite{stingray, liquidio, mellanox_bluefield}, FairNIC~\cite{fairnic} partitions cores, caches, and memory across tenants and shares bandwidth across tenants through Deficit Weighted Round Robin (DWRR) scheduling. iPipe~\cite{ipipe} uses a hybrid DRR+FCFS scheduler to share SmartNIC and host processors between different programs. \sysname uses space partitioning as well to allocate different resources to different modules. However, RMT's spatial/dataflow architecture differs considerably from the Von Neumann architectures for multi-core network processors targeted by FairNIC and iPipe. An RMT architecture can not support a runtime system similar to the ones used by iPipe and FairNIC.

\Para{FPGA-based solutions.} Several FPGA platforms exist for programmable packet processing. These platforms can be broadly categorized into (1) direct programming of FPGAs~\cite{clicknp, azure_smartnic, innova, p4netfpga, virtp4, mtpsa, p4fpga} and (2) higher-level abstractions built on top of FPGAs~\cite{flowblaze, hxdp, switchblade, nica}.

Systems (e.g., VirtP4~\cite{virtp4}, MTPSA~\cite{mtpsa}) based on direct FPGA programming typically implement packet-processing logic in a hardware-description language (HDL) or using a high-level language like P4~\cite{p4fpga, p4netfpga} or C~\cite{xilinx_hls, clicknp} that is translated into HDL. The HDL program is fed to an FPGA synthesis tool to produce a bitstream, which is written into the FPGA.
This approach requires combining the programs of different modules into a single Verilog program, which can then be fed to the synthesis tool. Thus, changing one module disrupts other modules, violating our requirement of no disruption.


FlowBlaze~\cite{flowblaze}, SwitchBlade~\cite{switchblade}, and hXDP~\cite{hxdp} expose a restricted higher-level abstraction like RMT or eBPF on top of an FPGA. FlowBlaze and hXDP do not provide support for isolation. SwitchBlade does, but its higher-level abstraction is much less flexible than the RMT abstraction in \sysname. NICA~\cite{nica} targets an FPGA NIC and is designed to share one pre-programmed offloading engine across many modules, while \sysname also targets ASIC pipelines and supports reprogramming individual modules without disrupting others.


\Para{Tofino~\cite{tofino}.} Tofino is a commercial switch ASIC that uses multiple parallel RMT pipelines. However, Tofino currently does not support multiple modules/P4 programs within a single pipeline. The current Tofino compiler requires a single P4 program per pipeline. Multiple P4 programs can be merged into a single program per pipeline and then fed into the Tofino compiler (Wang et al.~\cite{multitenant_hotcloud} and $\mu$P4~\cite{microp4}). However, both approaches still disrupt all tenants every time a single tenant in any pipeline is updated. This is because despite supporting an independent program per pipeline, updating any of these programs requires a reset of the entire Tofino switch~\cite{tofino_fast_refresh}. 



\Para{Emulation-based solutions.} Hyper4~\cite{hyper4} and HyperV~\cite{hyperv} propose to emulate multiple P4 programs/modules using a single hypervisor P4 program, which can be configured at run time by the control plane, thus supporting disruption-free reconfiguration. However, we found that it was very challenging to design a sufficiently ``universal'' hypervisor program on a commercial RMT switch like Tofino.

As one example, the hypervisor program needs to support performing a bit-shift by an amount determined by a packet field, where the packet field is specified by the control plane. However, a high-speed chip like Tofino has several restrictions on bit-shifts and other computations for performance, e.g., on Tofino, the shift width and field to shift must be supplied at compile time, not at run time by the control plane.

\Para{PANIC~\cite{panic} and FlexCore~\cite{flexcore}.} PANIC and FlexCore~\cite{flexcore} are programmable multi-tenant NIC and switch designs, respectively. They both suffer from scalability issues because they need to build a large crossbar with long wires interconnecting all engines to each other, which requires careful physical design~\cite[Appendix C]{drmt}. \sysname's RMT pipeline is easier to scale as its wires are shorter: they only connect adjacent pipeline stages~\cite[2.1]{rmt}.



\section{Conclusion}
\label{sec:conclusion}

This paper described \sysname, a system for isolating co-resident packet-processing modules on pipelines similar to RMT. \sysname builds on the idea of space partitioning and overlays, and is comprised of a set of simple hardware primitives that are inserted at different points in an RMT pipeline. These primitives are straightforward to realize in both ASICs and FPGAs. \sysname thus demonstrates that providing inter-module isolation in high-speed packet-processing pipelines is practical. Our software and hardware are available at \gitrepo.

\noindent\textbf{Acknowledgements.} We thank the NSDI reviewers and our shepherd Rodrigo Fonseca for their insightful comments and suggestions. We also thank Mike Walfish, Ravi Netravali, Mina Tahmasbi Arashloo, Amy Ousterhout, and Fabian Ruffy for their suggestions on this paper. We thank Han Wang and Anurag Agrawal with whom we discussed the Tofino architecture, and Alex Forencich, the FlowBlaze and NetFPGA teams, who helped us with debugging and design. This work was funded in part by NSF grants CCF-2028832, CNS-2008048, UK EPSRC project EP/T007206/1, and a gift from Google.

\clearpage
\balance
\bibliographystyle{abbrv}
\bibliography{ref}

\clearpage
\appendix
\section{Daisy-Chain vs. Fully-AXI-L-Based Configuration}
\label{app:daisy}
As discussed in \cref{subsec:hardware}, \sysname uses a daisy chain pipeline to configure the \sysname pipeline and uses the AXI-L~\cite{axil} protocol for safety alone, i.e., to read the reconfiguration packet counter and update the bitmap during reconfiguration. Before using this daisy-chain approach, we considered a different approach based fully on the AXI-L protocol. In this approach, all configuration settings on the FPGA would be set using the AXI-L protocol via PCIe from the host instead of passing a reconfiguration packet through a daisy chain pipeline. We elected to use the daisy-chain approach instead for 2 reasons described below.

\begin{figure}[h]
    \centering
    \includegraphics[width=0.9\linewidth]{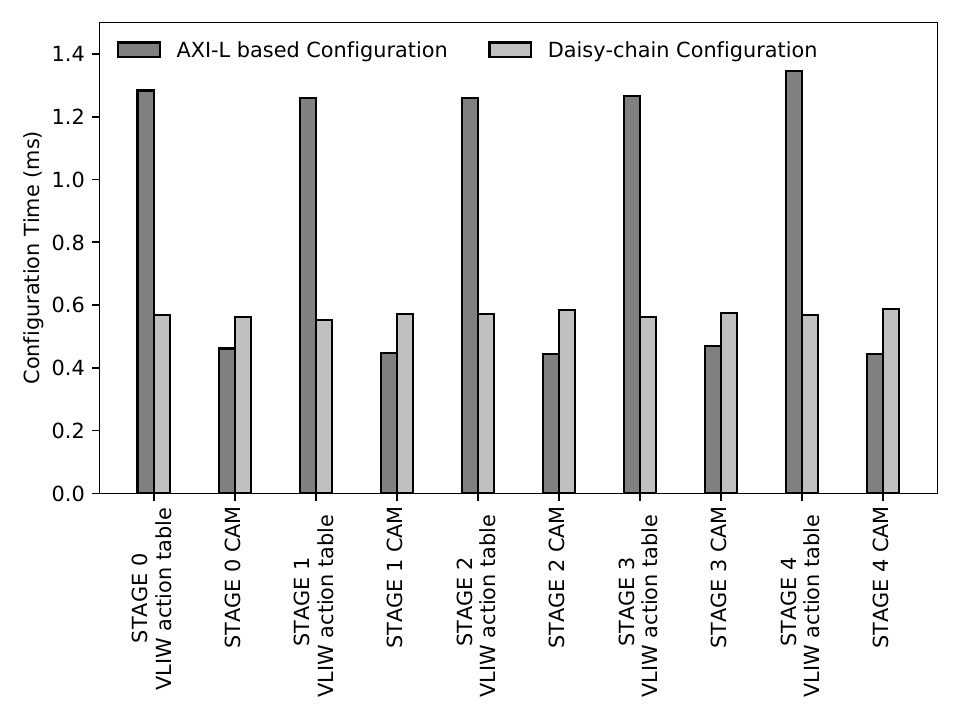}
    \caption{Configuration time comparison for AXI-L based (estimated) and \sysname's daisy-chain configuration (measured).}
    \label{exp_fig:axil_time}
\end{figure}

First, as one AXI-L write in Corundum can only support a 32-bit data length, we have to write $\lceil625/32\rceil=20$ and $\lceil205/32\rceil=7$ times for configuring one entry in the VLIW action table and CAM respectively. For our test modules, we estimate AXI-L reconfiguration time based on the write time of a single AXI-L write. As shown in Figure~\ref{exp_fig:axil_time}, \sysname's daisy-chain configuration is much faster than the AXI-L based method, especially for  longer entries (i.e., VLIW action table). These benefits are likely to be more pronounced on a larger implementation of \sysname because the entries (both for VLIW action table and CAM) will be even longer in that case. Second, the daisy-chain approach is more similar in style to how programmable switch ASICs are configured today, hence, it is preferable for an eventual ASIC implementation of \sysname.

\section{Isolation of ternary match tables using the Xilinx CAM IP}
\label{app:tcam}

While our current \sysname implementation only supports exact matching, we could reuse our implementation strategy (the Xilinx CAM IP) for ternary matching as well. However, supporting isolation between the ternary match tables of multiple different modules requires some care. This is to ensure that updates to the ternary match-action rules for one module do not cause updates to the ternary match-action rules for another module. 

In the case of ternary matching, the Xilinx CAM IP block uses the address of a CAM entry as the TCAM priority to determine which entry to return when there are multiple matches~\cite{xilinx_cam}. Concretely, the Xilinx CAM IP block can prioritize either the entry with the lowest address or the highest address. To support isolation on top of this block, first, we append the module ID (i.e., VLAN ID) to ternary match-action rules as we do currently for exact matches (\S\ref{sec:design}). Second, we allocate contiguous addresses within the Xilinx CAM IP block to a particular module.

Appending the module ID ensures that a module's packets do not match any other module's match-action rules. Allocating contiguous addresses ensures that a new match-action rule can be added (or an old rule can be updated) for a module with disruption to that module's match-action rules alone---and importantly, without disturbing the rules for any other modules.\footnote{Note that a new rule can be added to a module only if there are still empty addresses within that module's chunk of contiguously allocated addresses.}

\section{Hardware resources in \sysname}
\label{app:considered_hw_resources}

\begin{table}[btph]
\begin{center}
\scriptsize 
    \begin{tabular}{|c|l|}
        \hline
        \textbf{Hardware Resource}& \textbf{Description} \\
        \hline
        \hline
        \multirow{2}{*}{Packet Filter} & A 32-bit bitmap, \\
        &  and a 4-byte reconfiguration packet counter \\
        \hline
        \multirow{3}{*}{PHV} & 2-byte, 4-byte, 6-byte containers, \\
        & each type has 8 containers \\ 
        & a 32-byte container for platform-specific metadata \\
        \hline
        Parsing action & 16 bits wide \\
        \hline
        Parser and deparser table & 10 parsing actions, 160 bits wide, 32 entries deep \\
        \hline
        Key extractor table & 38 bits wide, 32 entries deep \\
        \hline
        Key mask table & 193 bits wide, 32 entries deep \\
        \hline
        Exact match table & 205 bits wide, 16 entries deep \\
        \hline
        ALU Action & 25 bits wide \\
        \hline
        VLIW action table & 25 ALU actions, 625 bits wide, 16 entries deep \\
        \hline
        Segment table & 16 bits wide, 32 entries deep \\
        \hline 
        Stages & 5 \\
        \hline
        Module ID & 12 bits \\
        \hline
    \end{tabular}
\caption{Hardware resources in \sysname}
\label{tab:parameters}
\end{center}
\end{table}

\section{Experimental setup}
\label{app:tb_setup}

\begin{figure}[!h]
    \centering
    \begin{subfigure}[t]{0.49\linewidth}
        \centering
        \includegraphics[width=\textwidth]{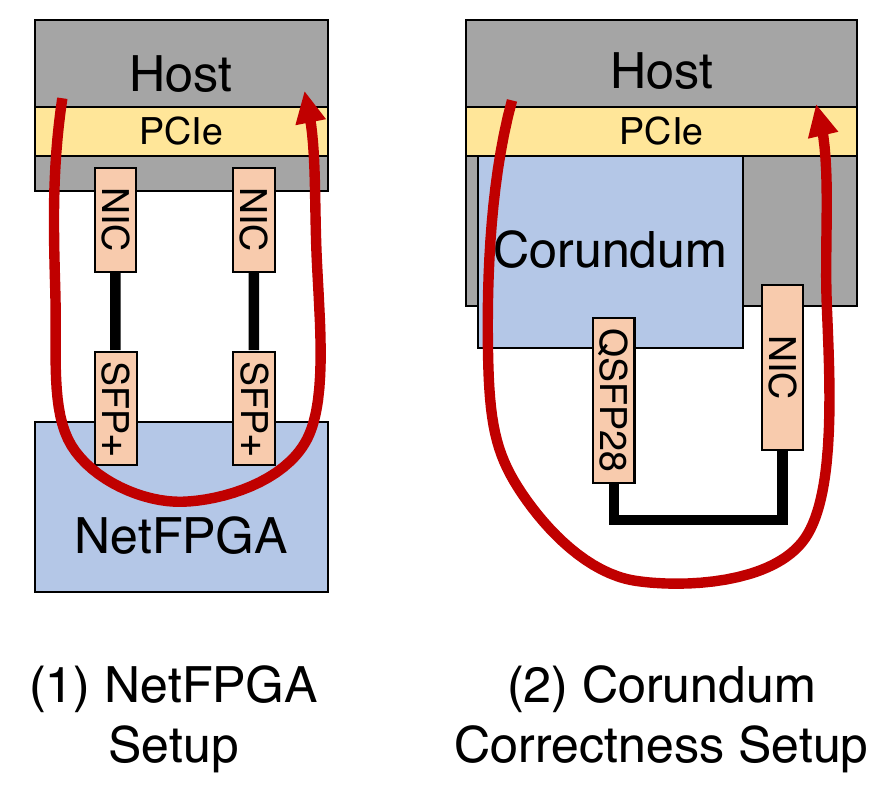}
        \caption{Correctness setup.}
        \label{fig:correctness_setup}
    \end{subfigure} \hfill
    \begin{subfigure}[t]{0.49\linewidth}
        \centering
        \includegraphics[width=0.63\textwidth]{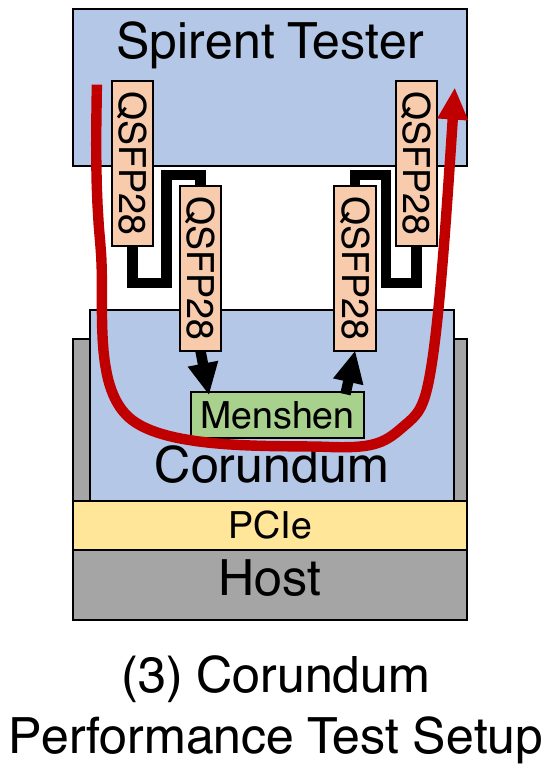}
        \caption{Performance test setup.}
        \label{fig:corundum_perf_setup}
    \end{subfigure} 
    \caption{Testbed setup. Red arrow shows packet flow.}
    \label{fig:exp_setup}
\end{figure}

\end{document}